\documentclass[useAMS, usenatbib]{mn2e}
\usepackage{color}

\usepackage{amsmath}
\usepackage{graphicx}

\newcommand{\ms}{\rm {h^{-1}M_{\odot}}}
\newcommand{\mpch}{{\rm h^{-1}Mpc}}

\begin{document}
\bibliographystyle{mn2e}
\graphicspath{{./figs/}}
\title[]{Measuring subhalo mass in redMaPPer clusters with CFHT Stripe 82 Survey}
\author[Ran Li et. al]
       {\parbox[t]{\textwidth}{
        Ran Li$^{1}$\thanks{E-mail:ranl@bao.ac.cn},
        Huanyuan Shan$^{2}$, 
        Jean-Paul Kneib$^{2,3}$,
        Houjun Mo$^{4}$,
        Eduardo Rozo$^{5}$,
        Alexie Leauthaud$^{6}$,
        John Moustakas$^{7}$,
        Lizhi Xie$^{8}$,
        Thomas Erben$^{9}$,
        Ludovic Van Waerbeke$^{10}$,
        Martin Makler$^{11}$,
        Eli Rykoff$^{12}$,
        Bruno Moraes$^{13,14}$      
       }
        \vspace*{3pt} \\
  $^{1}$Key laboratory for Computational Astrophysics, Partner Group of the Max Planck Institute for Astrophysics, \\
  National Astronomical Observatories, Chinese Academy of Sciences, Beijing, 100012, China\\
  $^{2}$Laboratoire d'Astrophysique, Ecole Polytechnique F\'ed\'erale de Lausanne (EPFL), 
          Observatoire de Sauverny, CH-1290 Versoix, Switzerland\\
  $^{3}$Aix Marseille Universit\'e, CNRS, LAM (Laboratoire d'Astrophysique de Marseille) UMR 7326, 13388, Marseille, France\\
  $^{4}$Department of Astronomy,  University of Massachusetts, Amherst MA 01003, USA \\
  $^{5}$Department of Physics, University of Arizona, 1118 E. Fourth St., Tucson, AZ 85721, U.S.A.\\
  $^{6}$Kavli Institute for the Physics and Mathematics of the Universe (Kavli IPMU, WPI), The University of Tokyo, Chiba 277-8583, Japan\\
  $^{7}$Department of Physics and Astronomy, Siena College, 515 Loudon Road, Loudonville, NY 12211, USA\\
  $^{8}$INAF, Astronomical Observatory of Trieste, Via Bazzoni 2, I-3424 Trieste, Italy\\
  $^{9}$Argelander-Institut f\"ur Astronomie Auf dem H\"ugel 71 D-53121 Bonn, German\\
  $^{10}$University of British Columbia, Department of Physics and Astronomy, 6224 Agricultural road, V6T 1Z1, Vancouver, Canada\\
  $^{11}$Centro Brasileiro de Pesquisas F\'isicas, Rua Dr Xavier Sigaud 150, CEP 22290-180, Rio de Janeiro, RJ, Brazil\\
  $^{12}$SLAC National Accelerator Laboratory, Menlo Park, CA 94025, U.S.A.\\
  $^{13}$Department of Physics and Astronomy, University College London, Gower Street, London, WC1E 6BT, UK\\
  $^{14}$CAPES Foundation, Ministry of Education of Brazil, Brasilia/DF 70040-020, Brazil\\
          }

\maketitle

\begin{abstract}
   We use the shear catalog from the CFHT Stripe-82 Survey to 
  measure the subhalo masses of satellite galaxies in redMaPPer clusters.
   Assuming a Chabrier Initial Mass Function (IMF) and a truncated NFW model
  for the subhalo mass distribution,  we find
  that the sub-halo mass to galaxy stellar mass ratio increases as a
    function of projected halo-centric radius $r_p$,
     from $M_{\rm sub}/M_{\rm star}=4.43^{+ 6.63}_{- 2.23}$ at $r_p \in [0.1,0.3]$ $\mpch$ to $M_{\rm sub}/M_{\rm star}=75.40^{+ 19.73}_{- 19.09}$
  at $r_p \in [0.6,0.9]$ $\mpch$. We also investigate the dependence of
  subhalo masses on stellar mass 
  by splitting satellite galaxies into two stellar mass bins: $10<\log(M_{\rm star}/\ms)<10.5$ and
   $11<\log(M_{\rm star}/\ms)<12$.  The best-fit subhalo mass of the more massive satellite galaxy
   bin is larger than that of the less massive satellites: $\log(M_{\rm sub}/\ms)=11.14 ^{+ 0.66 }_{- 0.73}$ ($M_{\rm sub}/M_{\rm star}=19.5^{+19.8}_{-17.9}$) versus $\log(M_{\rm sub}/\ms)=12.38 ^{+ 0.16 }_{- 0.16}$ ($M_{\rm sub}/M_{\rm star}=21.1^{+7.4}_{-7.7}$).
\end{abstract}

\section{Introduction}

In a cold dark matter (CDM) universe, dark matter haloes form hierarchically through accretion and merging(for a recent review, see \citet{Frenk2012}). 
Many rigorous and reliable predictions for the halo mass function and subhalo mass functions in CDM are provided by numerical simulations\citep[e.g.][]{springel2009, Gao2011, Colin2000, Hellwing2015, Bose2016}.  When the small haloes merge to the larger systems, they become the subhaloes and suffer from environmental effects such as tidal stripping, tidal heating and dynamical friction that tend to remove the mass from them and even disrupt them \citep{ Tormen1998, Taffoni2003, Diemand2007, Hayashi2003, gao2004, springel2009, xie2015} .  At the mean time, the satellite
galaxies that reside in the subhaloes also experience environmental effects. The tidal stripping and ram-pressure can remove the hot 
gas halo from satellite galaxies which in turn cuts off their supply of cold gas and quenches star formation\citep{Balogh2000,Kawata2008,McCarthy2008,Wang2007,Guo2011,Wetzel2013}. Satellite galaxies in some cases also experience a mass loss in the cold gas component and stellar component during the interaction with the host haloes \citep{Gunn1972,Abadi1999,Chung2009, Mayer2001, Klimentowski2007, Kang2008,  Chang2013}. 

Overall, the subhaloes preferentially lose their dark mass rather than the luminous mass, because the mass distribution of satellite galaxies is much more concentrated than that of the dissipationless dark matter particles. Simulations predict that the mass loss of infalling subhaloes depends inversely on  their halo-centric radius \citep[e.g.][]{Springel2001, DeLucia2004, gao2004, xie2015}. Thus, the halo mass to stellar mass ratio of satellite galaxies should increase as a function of halo-centric radius.  Mapping the mass function of subhaloes from observations can provide important constraints on this galaxy evolution model.

In observations, dark matter distributions are best measured with gravitational lensing. For dark matter subhalos, however, such observations are 
challenging due to their relative low mass compared to that of the host dark matter halo.
The presence of subhalos can cause flux-ratio anomalies in multiply imaged lensing systems \citep{Mao1998,Metcalf2001,Mao2004,xu2009,  Nierenberg2014}, perturb the locations, and change the image numbers \citep{kneib1996,Kneib2011}, and
 disturb the surface brightness of extended Einstein ring/arcs  \citep{Koopmans2005,Vegetti2009a,Vegetti2009b,Vegetti2010,Vegetti2012}.
However, due to the limited number of high quality images and the rareness of strong lensing systems, only a few subhalos have been detected
through strong lensing observation. Besides, strong lensing effects can only probe the central regions of dark matter haloes \citep{Kneib2011}. 
Therefore, through strong lensing alone, it is difficult to draw a
comprehensive picture of the co-evolution between subhalos and galaxies.

Subhalos can also be detected in individual clusters
through weak gravitational lensing or weak lensing combining strong lensing
 \citep[e.g.][]{Natarajan2007,Natarajan2009, Limousin2005,Limousin2007,Okabe2014}. 
In \citet{Natarajan2009},  Hubble Space Telescope images were used to investigate the
 subhalo masses of $L^*$ galaxies in the massive, lensing cluster,
 Cl0024+16 at z = 0.39, and to study the subhalo
  mass as a function of halo-centric radius. \citet{Okabe2014} investigated subhalos
 in the very nearby Coma cluster with imaging from the Subaru telescope. The deep imaging and the large apparent size of the
cluster allowed them to measure the masses of subhalos selected by
shear alone. They found 32 subhalos in the Coma cluster 
and measured their mass function. However, this kind of study requires very high quality images of  massive nearby clusters,
making it hard to extend such studies to large numbers of clusters. 

A promising alternative way to investigate the satellite-subhalo
relation is through a stacking analysis of galaxy-galaxy lensing with
large surveys. Different methods have been proposed in previous studies \citep[e.g.][]{Yang2006, Li2013,Pastor2011,Shirasaki2015}. 
Although the tangential shear generated by a single subhalo is small,
by stacking thousands of satellite galaxies the statistical noise can
be suppressed and the mean projected density
profile around satellite galaxies can be measured. \citet{Li2014} selected
satellite galaxies in the SDSS group catalog from \citep{Yang2007} and measured the weak lensing signal
around these satellites with a lensing source catalog derived from the CFHT Stripe82 Survey \citep{Comparat2013}. This was the first galaxy-galaxy lensing
 measurement of subhalo masses in galaxy groups. However, the uncertainties of the measured subhalo
masses were too large to investigate the satellite-subhalo relation as a function of halo-centric radius.

In this paper, we apply the same method as  \citet{Li2013,Li2014} to measure the galaxy-galaxy lensing signal
for  satellite galaxies in the SDSS redMaPPer cluster catalog \citep{Rykoff2014, Rozo2014}. Unlike the
the group catalog of \citet{Yang2007} which is constructed using SDSS spectroscopic galaxies, the redMaPPer
catalog relies on photometric cluster detections, allowing it to go to higher redshifts.  As a result,
there are more massive clusters in the redMaPPer cluster
catalog. Therefore, we expect to signal-to-noise of the satellite galaxy lensing signals
to be higher, enabling us to derive better constraints on subhalo properties.

The paper is organized as follows. In section \ref{sec:data}, we describe the lens and
source catalogs. In section \ref{sec:model}, we present our lens model. In section \ref{sec:res}, 
we show our observational results and our best fit lens model. 
The discussions and conclusions are presented in section \ref{sec:sum}. Throughout the paper, we adopt a $\Lambda$CDM cosmology with parameters given by the WMAP-7-year data \citep{komatsu2010} with $\Omega_L=0.728$, $\Omega_{M}=0.272$ and $h \equiv H_0/(100 {\rm km s^{-1} Mpc^{-1}}) = 0.73$. In this paper, stellar mass is estimated assuming a \cite{Chabrier2003} IMF.

\section{Observational Data}
\label{sec:data}

\subsection{Lens selection and stellar masses}
\label{sec:lens_select}

We use satellite galaxies in the redMaPPer clusters as lenses. The redMaPPer cluster
 catalog  is extracted from photometric galaxy samples of the SDSS
  Data Release 8 \citep[DR8,][]{Aihara2011} using the red-sequence Matched-filter
   Probabilistic Percolation cluster
 finding algorithm \citep{Rykoff2014}. The redMaPPer algorithm uses the 5-band
$(ugriz)$  magnitudes of galaxies with a magnitude cut $i<21.0$ over a total area of 10,000 deg$^2$
to photometrically detect galaxy clusters.

redMaPPer uses a multi-color richness estimator $\lambda$,
 defined to be the sum of the membership probabilities over all galaxies. 
 In this work,  we  use clusters with richness $\lambda>20 $ and photometric redshift $z_{\rm cluster}<0.5$. 
  In the overlapping region with the CFHT Stripe-82 Survey,  we have a total of 634 clusters. For each redMaPPer cluster, 
 member galaxies are identified according to their
 photometric redshift,  color and their cluster-centric distance.  To reduce the contamination
 induced by fake member galaxies, we only use satellite galaxies with membership probability $P_{\rm mem}>0.8$.  
 The redMaPPer cluster finder identifies 5  central galaxy candidates for each cluster, each with an estimate of the probability $P_{\rm cen}$
 that the galaxy in question is the central galaxy of the cluster.
We remove all central galaxy candidates  from our lens sample. For more details  about
redMaPPer cluster catalog, we refer the readers to \citet{Rykoff2014, Rozo2014}. 

\begin{figure}
\includegraphics[width=0.4\textwidth]{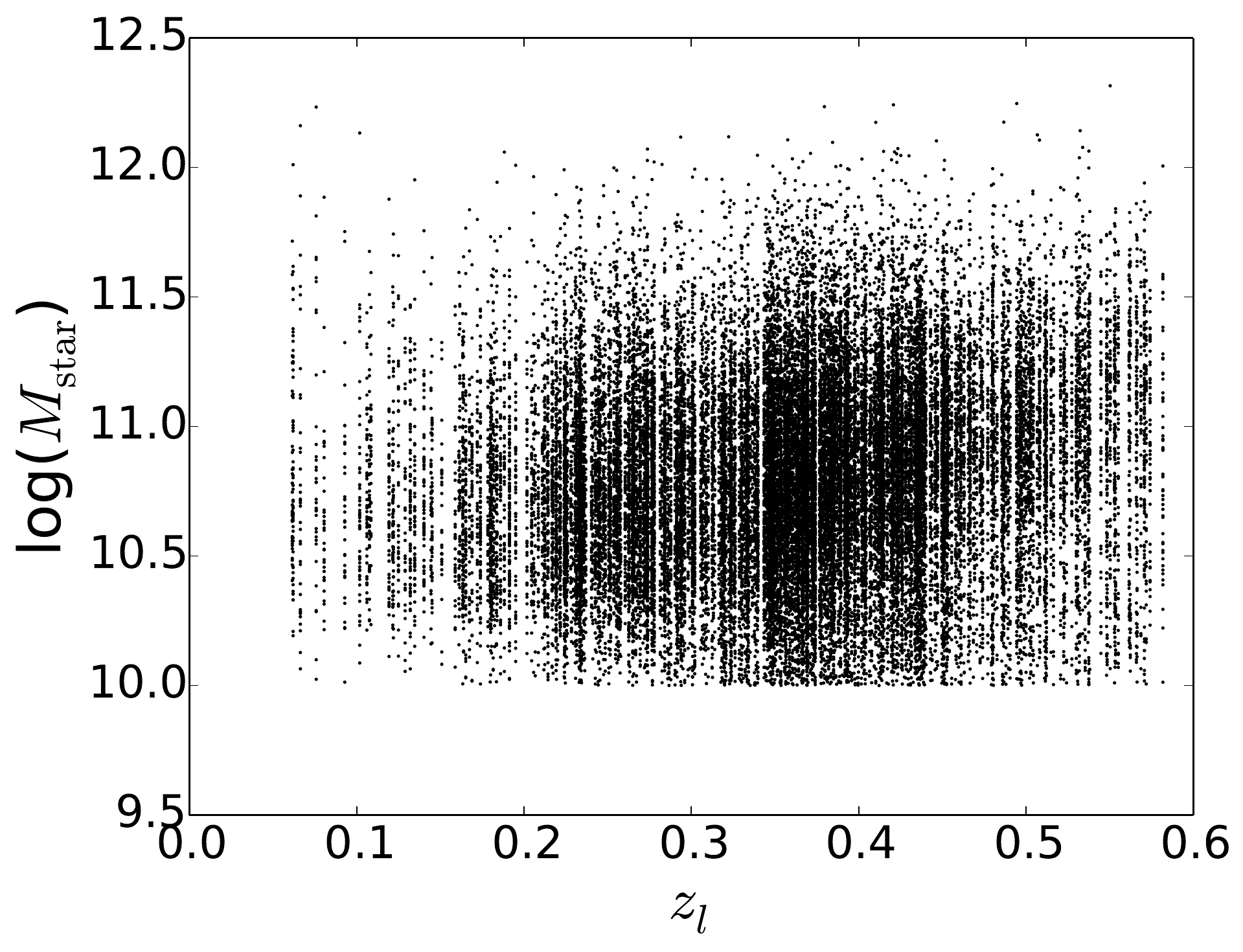}
\caption{The $M_{\rm star}$--$z_l$ distribution of lens galaxies, where
$z_l$ is the photometric redshift, and $M_{\rm star}$ is in units of $M_{\odot}$. }
\label{fig:mz}
\end{figure}

Stellar masses are estimated for member galaxies in the redMaPPer
catalog using the Bayesian spectral energy distribution (SED) modeling
code {\tt iSEDfit} \citep{Moustakas2013}.  Briefly, {\tt iSEDfit}
determines the posterior probability distribution of the stellar mass
of each object by marginalizing over the star formation history,
stellar metallicity, dust content, and other physical parameters which
influence the observed optical/near-infrared SED.  The input data for
each galaxy includes: the SDSS $ugriz$ {\tt model} fluxes scaled to the
$r$-band {\tt cmodel} flux; the 3.4- and 4.6-$\micron$ ``forced" WISE
\citep{Wright2010} photometry from \citet{Lang2014}; and the
spectroscopic or photometric redshift for each object inferred from
redMaPPer.  We adopt delayed, exponentially declining star formation
histories with random bursts of star formation superposed, the
Flexible Stellar Population Synthesis (FSPS) model predictions of
\citet{Conroy2009, Conroy2010}, and the \citet{Chabrier2003} initial
mass function (IMF) from 0.1-100 $M_{\odot}$.  For reference, adopting
the \citet{Salpeter1955} IMF would yield stellar masses which are on
average 0.25 dex (a factor of 1.8) larger. We apply a
  stellar mass cut of $M_{\rm star}>10^{10} \ms$ to our satellite
  galaxy sample. In Fig.\ref{fig:mz}, we show the $M_{\rm
    star}$--$z_l$ distribution for our lens samples, where $z_l$ is
  the photometric redshift of the satellite galaxy assigned
  by the redMaPPer algorithm.  The low stellar mass satellite galaxies are
   incomplete at higher redshift, but they will not affect the conclusion of this 
   paper. 

\subsection{The Source Catalog}

The source catalog is measured from the Canada--France--Hawaii
Telescope Stripe 82 Survey (CS82),  which is an $i-$band imaging survey
covering the SDSS Stripe82 region. With excellent seeing conditions --- FWHM between 0.4 to 0.8 arcsec ---
the CS82 survey reaches a depth of $i_{\rm AB}\sim24.0$.
The survey contains a total of 173 tiles, 165 of which from CS82 observations and 8 from CFHT-LS Wide \citep{Erben2013}.
The CS82 fields were observed in four dithered observations with 410s
exposure.
The 5$\sigma$ limiting magnitude is $i_{\rm AB}\sim 24.0$ in a 2 arcsec diameter aperture.

Each CFHTLenS science image is supplemented by a mask,
indicating regions within which accurate photometry/shape measurements
of faint sources cannot be performed, e.g. due to extended
haloes from bright stars. According to \citet{Erben2013}, most of the science analysis are safe with
sources with MASK$\le1$. After applying all the necessary masks and removing overlapping regions,
 the effective survey area reduces from 173 deg$^2$ to 
129.2 deg$^2$.  We also require that source galaxies to have FITCLASS= 0, where FITCLASS is the flag that
describes the star/galaxy classification.

Source galaxy shapes are measured with the lensfit method \citep{Miller2007,  Miller2013}, 
closely following the procedure in \citet{ Erben2009,  Erben2013}. The
shear calibration and systematics of the
lensfit pipeline are described in detail in \citet[][]{Heymans2012}. The specific procedures that 
are applied to the CS82 imaging are described in  Erben et al. (2015,  in preparation).

Since the CS82 survey only provides the $i$-band images, the CS82 collaboration derived  source photometric redshifts
using the $ugriz$ multi-color data from the SDSS co-add \citep{Annis2014}, which reaches roughly 2
magnitudes deeper than the single epoch SDSS imaging. The photometric redshift (photo-z) of the background
galaxies were estimated using a Bayesian photo-$z$ code \citep[BPZ,][]{Benitez2000, Bundy2015}. 
The effective weighted source galaxy number density is 4.5 per arcmin$^{2}$. Detailed systematic tests for this weak lensing
catalog are described in Leauthaud et al. 2015 (in prep).

\subsection{Lensing Signal Computation}

In a galaxy-galaxy lensing analysis, the excess surface mass density
,  $\Delta\Sigma$ is inferred  by measuring the tangential shear,  $\gamma_t(R)$,  where
\begin{equation}\label{eq:ggl}
\Delta\Sigma(R)=\gamma_t(R)\Sigma_{\rm crit}={\overline\Sigma}(<R)-\Sigma(R)\, , 
\end{equation}
where ${\overline\Sigma}(<R)$ is the mean surface mass density
within $R$,  and $\Sigma(R)$ is the average surface density at the projected radius $R$,  and 
$\Sigma_{\rm crit}$ is the lensing critical surface density
\begin{equation}
  \Sigma_{\rm crit}=\frac{c^2}{4\pi G}\frac{D_s}{D_l D_{ls}}\, , 
\end{equation}
where $D_{ls}$ is the angular diameter distance between the lens and the source,
and $D_l$ and $D_s$
are the angular diameter distances from the observer to the lens and
to the source,  respectively.

We select satellite galaxies as lenses and stack
lens-source pairs in physical radial distance $R$ bins from $0.04$ to $1.5 \mpch$.
To avoid contamination from foreground galaxies, we remove lens-source pairs 
with $z_{s}-z_{l} < 0.1$, where $z_s$ and $z_l$ are lens redshift and source redshift respectively.
We have also tested the robustness of our results by varying the selection criteria for source galaxies.
We find that selecting lens-source pairs with $z_{s}-z_{l} > 0.05$ or $z_{s}-z_{l} > 0.15$ only changes the
final lensing signal by less than 7\%, well below our final errors.

For a given set of lenses,  $\Delta\Sigma(R)$ is estimated using
\begin{equation}
\Delta\Sigma(R)=\frac{\sum_{l}\sum_{s}w_{ls}\gamma_t^{ls}\Sigma_{\rm crit}(z_l,z_s)}{\sum_{l}\sum_{s}w_{ls}}\, , 
\end{equation}
where
\begin{equation}
w_{ls}=w_n\Sigma_{\rm crit}^{-2}(z_l,z_s)\, 
\end{equation}
and $w_n$  is a weight factor defined by Eq.\, (8) in \citet{Miller2013}. 
$w_n$ is introduced to account for
 the intrinsic distribution of ellipticity and shape measurement uncertainties.

In the lensfit pipeline, a calibration factor for the multiplicative error $m$ is estimated
for each galaxy based on its signal-to-noise ratio, and the size of the galaxy. 
Following \citet{Miller2013}, we account for these multiplicative errors 
in the stacked lensing by the correction factor
\begin{equation}
1+K(R)=\frac{\sum_{l}\sum_{s} w_{ls}(1+m_s)}{\sum_{l}\sum_{s} w_{ls}}
\end{equation}

The corrected lensing measurement is as:
\begin{equation}
\Delta\Sigma(R)^{\rm corrected}=\frac{\Delta\Sigma(R)}{1+K(R)}
\end{equation}

In this paper, we stack the tangential shears around satellite galaxies binned according to their
projected halo-centric distance $r_p$, and fit the galaxy-galaxy lensing signal to
obtain the subhalo mass of the satellite galaxies. We describe our theoretical lens
models below.

\section{The Lens Model}
\label{sec:model}

The surface density around a lens galaxy  $\Sigma(R)$ can be written as: 
\begin{equation}\label{xi_gm}
\Sigma(R)=\int_{0}^{\infty}\rho_{\rm g, m} \left(\sqrt{R^2+\chi^2} \right)\, {\rm d}\chi\, ;
\end{equation}
and the mean surface density within radius  $R$ is
\begin{equation}\label{xi_gm_in}
\Sigma(< R)=\frac{2}{R^2} \int_0^{R} \Sigma(u) \,  u \,  {\rm d}u , 
\end{equation}
where $\rho_{g, m}$ is the density profile around the lens,  and 
$\chi$ is the physical distance along the line of sight.  
The excess surface density around a satellite galaxy is composed of three
components:
\begin{equation}
\Delta\Sigma(R)=\Delta\Sigma_{\rm sub}(R)+\Delta\Sigma_{\rm host}(R,  r_p)+\Delta\Sigma_{\rm star}(R)\, , 
\end{equation}
where $\Delta\Sigma_{\rm sub}(R)$ is the contribution of the subhalo in which the satellite 
galaxy resides,  $\Delta\Sigma_{\rm host}(R,  r_p)$ is the contribution from the host halo of the cluster/group, 
where $r_p$ is the  projected distance from the satellite galaxy  to the center of the host halo,
and $\Delta\Sigma_{\rm star}(R)$ is the contribution from 
  the stellar component of the satellite galaxy. We neglect the two-halo term, which is the contribution from 
  other haloes on the line-of-sight, because this contribution is only
  important
 at $R>3 \mpch$ for clusters \citep[see][]{Shan2015}.
  At small scales where the subhalo term dominates, the contribution
  of the 2-halo term is at least an order of magnitude
  smaller than the subhalo term \citep{Li2009}.

\subsection{Host halo contribution}

We assume that  host haloes are centered on the central galaxies of clusters, 
with a density profile following the NFW \citep{NFW97} formula:
\begin{equation}
\rho_{\rm host}(r)=\frac{\rho_{\rm 0,host}}{(1+r/r_{\rm s,host})^2(r/r_{\rm s,host})} \,,
\end{equation}
where $r_{\rm s,host}$ is the characteristic scale of the halo and $\rho_{\rm 0,host}$ is a normalization 
factor.  Given the mass of a dark matter halo, its profile then only depends on the concentration
parameter $c\equiv r_{\rm 200}/r_{\rm s,host}$, where $ r_{\rm 200}$ is a radius within which the average
density of the halo equals to 200 times the universe critical density, $\rho_{\rm crit}$.
The halo mass $M_{\rm 200}$ is defined as $M_{\rm 200}\equiv 4\pi/3r_{\rm 200}^3(200 \rho_{\rm crit})$.

Various fitting formulae for mass-concentration relations have been derived from N-body numerical simulations
  \citep[e.g.][]{Bullock2001,Zhao2003,Dolag2004,Maccio2007,Zhao2009, Duffy2008, Neto2007}.
 These studies find that the concentration decreases with increasing
 halo mass. Weak lensing observations also measure this trend,
 but the measured amplitude of the mass-concentration relation is  slightly smaller than that in the simulation
 \citep[e.g.][]{Mandelbaum2008, Miyatake2013, Shan2015}.
Since there is almost no degeneracy between  the subhalo mass and the
concentration \citep{Li2013,George2012},
we expect that the exact choice of the mass-concentration relation
should not have a large impact on our conclusions.
 Throughout this paper,  we adopt the mass concentration relation from
 \citet{Neto2007}:
 \begin{equation}
 c=4.67(M_{\rm 200}/10^{14} \ms)^{-0.11}
 \end{equation}

We stack satellite galaxies with different halo-centric distance  $r_p$, and in host halos
with different mass. Thus the lensing contribution from the 
host halo is an average of $\Delta{\Sigma}_{\rm host}$ over $r_p$ and 
host halo mass $M_{\rm 200}$. The host halo profile is modeled as follows.

For each cluster, we can estimate its mass via the richness-mass relation of \citet{Rykoff2012}:
\begin{equation}
\label{eq:mass_rich}
  \ln{\left( \frac{M_{\rm 200}}{h_{\rm 70}^{-1}10^{14}M_{\odot}} \right)} = 1.48 + 1.06\ln(\lambda/60)
\end{equation}
In the redMaPPer catalog,  each cluster has five possible central galaxies, each with probability $P_{\rm cen}$.
We assume that the average $\Delta\Sigma$ contribution from the host halo to a satellite can be written as:
\begin{equation}
  \Delta\bar{\Sigma}_{\rm host}(R)=A_0 \sum_{i}^{N_{\rm sat}}\sum_{j}^5\Delta\Sigma_{\rm host}(R|r_{\rm p, j},  M_{\rm 200}) P_{\rm cen, j}\, , 
\end{equation}
where $r_{\rm p, j}$ is the projected distance between the satellite and the jth candidate of the host cluster center,  $P_{\rm cen, j}$
is the probability of jth candidate to be the true center, and $N_{\rm
  sat}$  is the number of stacked satellite galaxies. $A_0$ is
the only free parameter of the host halo contribution model. It describes an adjustment of the lensing amplitude. If the richness-mass relation
is perfect, the best-fit $A_0$ should be close to unity.  Note that, the subhalo mass determination is robust against the variation of 
the normalization in richness-mass relation. If we decrease the normalization in Eq. \ref{eq:mass_rich} by 20\%, the best-fit subhalo mass
changes only by 0.01 dex, which is at least 15 times smaller than the $1\sigma$ uncertainties of $M_{\rm sub}$ (see table \ref{tab:para_nfw}).

\subsection{Satellite contribution}

In  numerical simulations, subhalo density profiles are found to 
be truncated in the outskirts \citep[e.g.][]{Hayashi2003,springel2009, gao2004, xie2015}.
In this work, we use a truncated NFW profile \citep[][tNFW, hereafter]{Baltz2009, Oguri2011}
to describe the subhalo mass distribution:
\begin{equation}\label{eq:rhosub} 
\rho_{\rm sub}(r)=\frac{\rho_{0}}{r/r_s(1+r/r_s)^2}\left(\frac{r_t^2}{r^2+r_t^2}\right)^2 \,
\end{equation}
where $r_t$ is the truncation radius of the subhalo, $r_s$ is the characteristic radius of the tNFW profile
and $\rho_0$ is the normalization.  The enclosed mass with $x\equiv r/r_s$ can be written as:%

\begin{eqnarray}
M(<x)&=&4\pi\rho_0 r_s^3 \frac{\tau^2}{2(\tau^2+1)^3(1+x)(\tau^2+x^2)}\nonumber\\
&&\hspace*{-16mm}\times\Bigl[(\tau^2+1)x\left\{x(x+1)-\tau^2(x-1)(2+3x)
-2\tau^4\right\}\nonumber\\
&&\hspace*{-16mm}+\tau(x+1)(\tau^2+x^2)\left\{2(3\tau^2-1)
{\rm arctan}(x/\tau)\right.\nonumber\\
&&\hspace*{-16mm}\left.+\tau(\tau^2-3)\ln(\tau^2(1+x)^2/(\tau^2+x^2))\right\} \Bigr],
\label{eq:mbmo_nodim}
\end{eqnarray}
where $\tau\equiv r_t/r_s$. We define the subhalo mass $M_{\rm sub}$ to be the total mass within
$r_t$. Given $M_{\rm sub}$, $r_s$ and $r_t$, the tangential  shear $\gamma_t$ can be derived analytically
(see the appendix in \citet{Oguri2011}).

Previous studies have sometimes used instead the pseudo-isothermal elliptical mass
distribution (PIEMD) model derived by \citet{Kassiola1993}) for modeling the mass distribution
around galaxies \citep[e.g.][]{Limousin2009,Natarajan2009,Kneib2011}. 
The surface density of the PIEMD model can be written as:
\begin{equation}\label{eq:PIEMD}
\Sigma(R)=\frac{\Sigma_0 R_0}{1-R_0/R_t}\left(\frac{1}{\sqrt{R_0^2+R^2}} - \frac{1}{\sqrt{R_t^2+R^2}} \right)\,
\end{equation}
where $R_0$ is the  core radius, $R_t$ is the truncation radius, and $\Sigma_0$ is a characteristic surface density.
The subhalo mass $M_{\rm sub}$, which is defined to be the enclosed mass with $R_t$, can be written as:
\begin{equation}
M_{\rm sub}=\frac{2\pi \Sigma_0 R_0 R_t}{R_t-R_0}\left[ \sqrt{R_0^2-R^2} - \sqrt{R_t^2-R^2}+(R_t-R_0)\right]
\end{equation}
In this paper, we will fit the data with both tNFW and PIEMD models.

Finally, the lensing contributed from stellar component is usually modeled as a point mass:
\begin{equation}
\Delta\Sigma_{\rm star}(R)=\frac{ M_{\rm star}}{R^2}\,,
\end{equation}
where $M_{\rm star}$ is the total stellar mass of the galaxy.

\section{Results}
\label{sec:res}
\subsection{Dependence on the projected halo-centric radius}

To derive the subhalo parameters, we calculate the $\chi^2$ defined as:
\begin{equation}
 \chi^{2}=\sum_{ij}({\Delta\Sigma(R_i)-\Delta\Sigma(R_i)^{obs}})(\widehat{C_{ij}^{-1}})({\Delta\Sigma(R_j)-\Delta\Sigma(R_j)^{obs}} )\,,
\label{chi2}
\end{equation}
where $\Delta\Sigma(R_i)$ and $\Delta\Sigma(R_i)^{obs}$ are the model and  the observed lensing signal in the $i$th radial bin.
The matrix $C_{ij}$ is the covariance matrix of data error between different radius bins. Even if the ellipticity of different sources are independent,
the cross term of the covariance matrix could still be non-zero, due to some source galaxies are used more than once\citep[e.g.][]{Han2015}.
The covariance matrix can be reasonably calculated with bootstrap method using the survey data themselves\citep{Mandelbaum2005}.
In the paper, We divide the CS82 survey area into 45 equal area sub-regions. We then generate 3000 bootstrap samples by resampling the 45 sub-regions of CS82 observation data sets and calculate the covariance matrix using the bootstrap sample. Thus, the likelihood function can be given as:
\begin{equation}
 L\propto \exp\left(-\frac{1}{2}\chi^2\right).
\label{likelihood}
\end{equation}

We select satellite galaxies as described in section \ref{sec:lens_select} 
and measure the stacked lensing signal around satellites in three
$r_p$ bins: $0.1<r_p<0.3$; $0.3<r_p<0.6$ and $0.6<r_p<0.9$(in unit of $\mpch$).

For each bin, we fit the stacked lensing signal with a
Monte-Carlo-Markov-Chain(MCMC) method. For the tNFW subhalo case, 
we have 4 free parameters:  $M_{\rm sub}$, the subhalo mass;  $r_s$, the tNFW profile
scale radius; $r_t$, the tNFW truncation radius, and $A_0$, the normalization factor
 of the host halo lensing contribution.
  
We adopt flat priors with broad boundaries for these model parameters. 
We set the upper boundaries for  $r_t$ and $r_s$ to be the value of the viral radius and
the scale radius of an NFW halo of $10^{13}\ms$.  We choose these values because 
the subhalo masses of satellite galaxy in clusters are expected to be much smaller than 
$10^{13}\ms$  \citep{Gao2012}.
 
For the PIEMD case, we also have 4 free
parameters: $M_{\rm sub}$, $R_0$, $R_t$ and $A_0$. We set the upper boundary of
 $R_t$ to be the same as $r_t$ in the tNFW case. We set the upper boundary of $R_0$ to
  be 20 kpc/h, which is much higher than the typical size of $R_0$  in
  observations\citep{Limousin2005, Natarajan2009}.  We believe our choice of priors is very conservative. 
  The detailed choices of priors are listed in table~\ref{tab:prior}. 

\begin{table}
\begin{center}
  \caption{Flat priors for model parameters. $M_{\rm sub}$
  is in units of $\ms$ and distances are in units of $\mpch$.}

\begin{tabular}{|c|c|c|}
 \hline   
 &&\\
 &lower bound& upper bound \\
   &&\\
  \hline 
 $A_{\rm 0}$          & 0.3& 2 \\
  \hline 
 $\log{M_{\rm sub}} $ & 9 & 13 \\
  \hline 
 $r_t$ (tNFW)         & 0 & 0.35 \\
  \hline 
 $r_s$ (tNFW)        & 0& 0.06 \\
  \hline 
 $R_0$(PIEMD)        & 0 & 0.02\\
  \hline 
 $R_t$(PIEMD)        & 0 & 0.35\\
  \hline 
\end{tabular}
\label{tab:prior}
\end{center}
\end{table}

\begin{figure}
\includegraphics[width=0.4\textwidth]{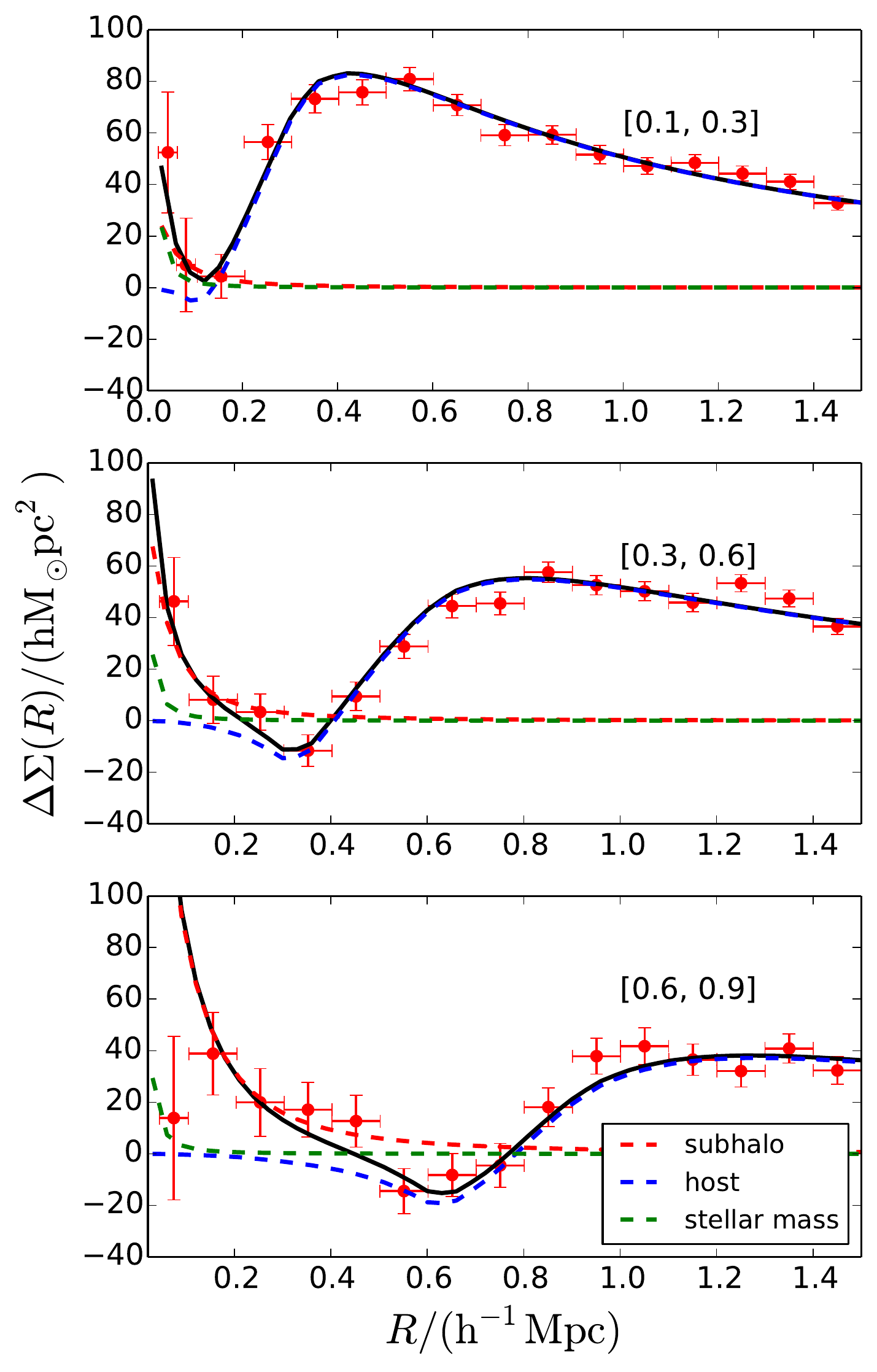}
\caption{Observed galaxy-galaxy lensing signal for satellite galaxies as a
function of the radius. The top,  middle and bottom panels show results
for satellites with $r_p$ in the range
$[0.1, 0.3]\mpch$,  $[0.3, 0.6]\mpch$ and $[0.6, 0.9]\mpch$, respectively. Red points with error bars
 represent the observational data. The vertical error bars show the
1 $\sigma$ bootstrap error. The horizontal error bars show the bin size. Black solid lines 
show the best-fit tNFW model prediction. Dashed lines of different colors show the contribution
from the subhalo, the host halo, and the stellar mass respectively.}
\label{fig:gglensing}
\end{figure}

In Fig.\ref{fig:gglensing}, we show the observed galaxy-galaxy lensing
signal.  Red dots with error bars represent the observational
data. The vertical error bars show 1 $\sigma$ errors estimated with
the bootstrap resampling the lens galaxies. 
Horizontal error bars show the range of each
radial bin. The lensing signals show clearly the characteristic shape
described in figure 2 of \citet{Li2013}.  The lensing signal from the
subhalo term  dominates the central part. On the other hand, the contribution from
the host halo is nearly zero on small scale, and decreases to negative values at intermediate
scale.   This is because  $\Sigma_{\rm host}(R)$ becomes increasingly larger compared to $\Sigma_{\rm host}(<R)$
at intermediate R.  At radii where $R>r_p$, $\Delta\Sigma_{\rm host}(R)$ increases  with R again. At large
scales where $R>>r_p$, $\Delta\Sigma(R,r_p)$ approaches
$\Delta\Sigma_{\rm host}(R,0)$. 

The solid lines show the best-fit results with the tNFW model. Dashed lines of different colors show the contribution from
different components. The best-fit model parameters are listed in
table~\ref{tab:para_nfw}. Note that, the value of the first point in
the $[0.6, 0.9]$ $r_p$ bin is very low, which may be due to the relatively
small number of source galaxies in the inner most radial bin. We exclude
this point when deriving our best fit model. For comparison, 
we also show the best-fit parameters including this point
in table~\ref{tab:para_nfw}.

\begin{figure*}
\includegraphics[width=\textwidth]{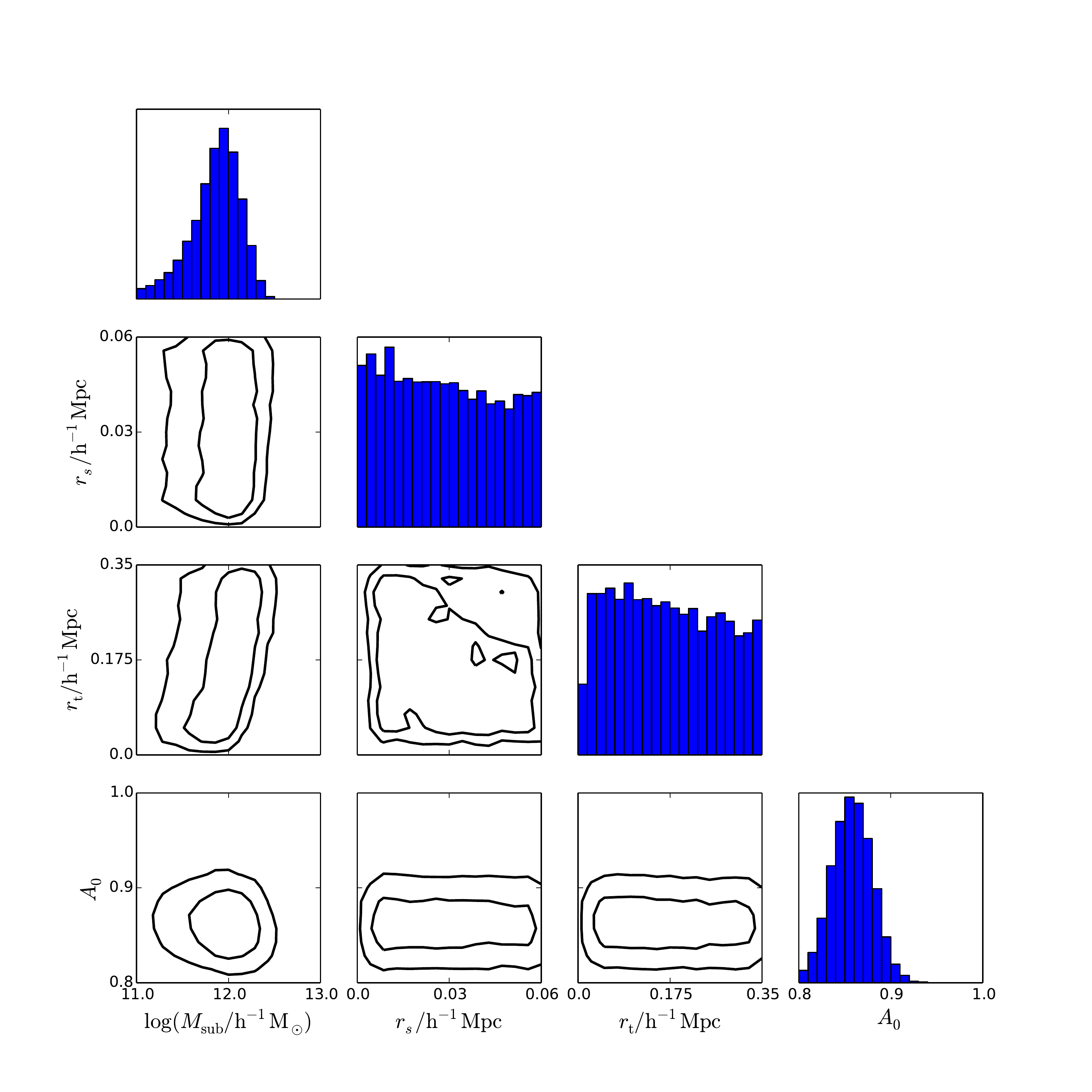}
\caption{The contours show 68\% and 95\% confidence intervals for the tNFW model parameters,
  $M_{\rm sub}$,  $r_s$,   $r_t$ and $A_0$. Results are shown for satellites with $r_p=[0.3, 0.6]\mpch$. 
  The last panel of each row shows the 1D marginalized posterior distributions.  Note that the plotting range
  for $r_s$ and $r_t$ is exactly the same as the limits of our prior, so we actually do not have much constraints
  for these two parameters, except that high values are slightly
  disfavored for both.  }
\label{fig:2dcontour}
\end{figure*}

Fig.~\ref{fig:2dcontour} shows an example of the MCMC  posterior
distribution of  the tNFW model parameters for satellites in the $[0.3, 0.6]\mpch$ $r_p$ bin.
The constraints on the subhalo mass $\log{M_{\rm sub}}$ are tighter than
that in \citet{Li2014}($\sim\pm 0.7$). The amplitude of $A_0$ is slightly smaller than unity, implying that
the clusters may be less massive than predicted by the mass-richness
relation. However, we are still unable to obtain significant 
constraints on the structure parameters of the sub-halos.
 In principle, the density profile cut-off caused by tidal effects
can be measured with tangential shear. However, the galaxy-galaxy lensing
measurement stacks many satellites, leading to smearing of the signal.  
With the data used here,  the tidal radius as a free parameter are not constrained.
Some galaxy-galaxy lensing investigations introduced additional 
constraints in order to estimate the tidal radius. For example, \citet{Gillis2013b}
assumed that galaxies of the same stellar mass but in different environments 
have similar sub-halo density profiles except the cut-off radius. 
With this additional assumption, they obtained $r_{\rm tidal}/r_{200}=0.26\pm0.14$. 
During the review process of our paper, \citet{Sifon2015} posted 
a similar galaxy-galaxy lensing measurement of satellite galaxies using 
the KiDs survey, and they also found that their data cannot distinguish 
models with or without tidal truncation.

\begin{figure}
\includegraphics[width=0.4\textwidth]{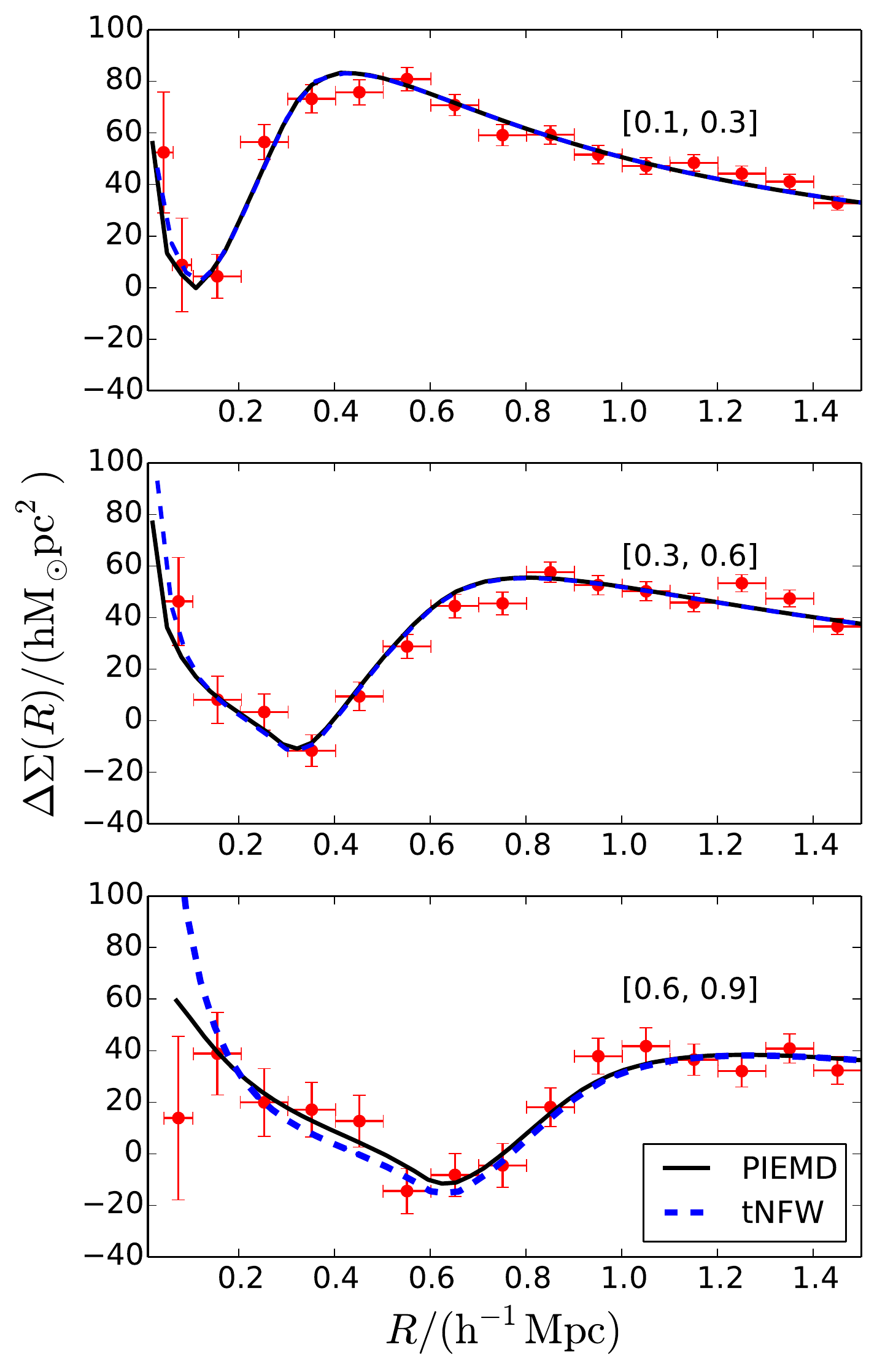}
\caption{This figure is similar to Fig\ref{fig:gglensing}. Solid lines represent the
    best-fit PIEMD model prediction and blue dashed lines are the
    predictions of the
        tNFW model.}
\label{fig:piemd}
\end{figure}

In Fig.\ref{fig:piemd}, we show the fitting results of the PIEMD model.
The best-fit parameters are listed in table~\ref{tab:para_piemd}.
For reference, we also over-plot the theoretical prediction of the best-fit
tNFW model with blue dashed lines. Both models provide a good description of the data. The best-fitted $M_{\rm sub}$ 
and $A_0$ of the two models agree well with each other, showing the validity
of our results.

\begin{figure}
\includegraphics[width=0.5\textwidth]{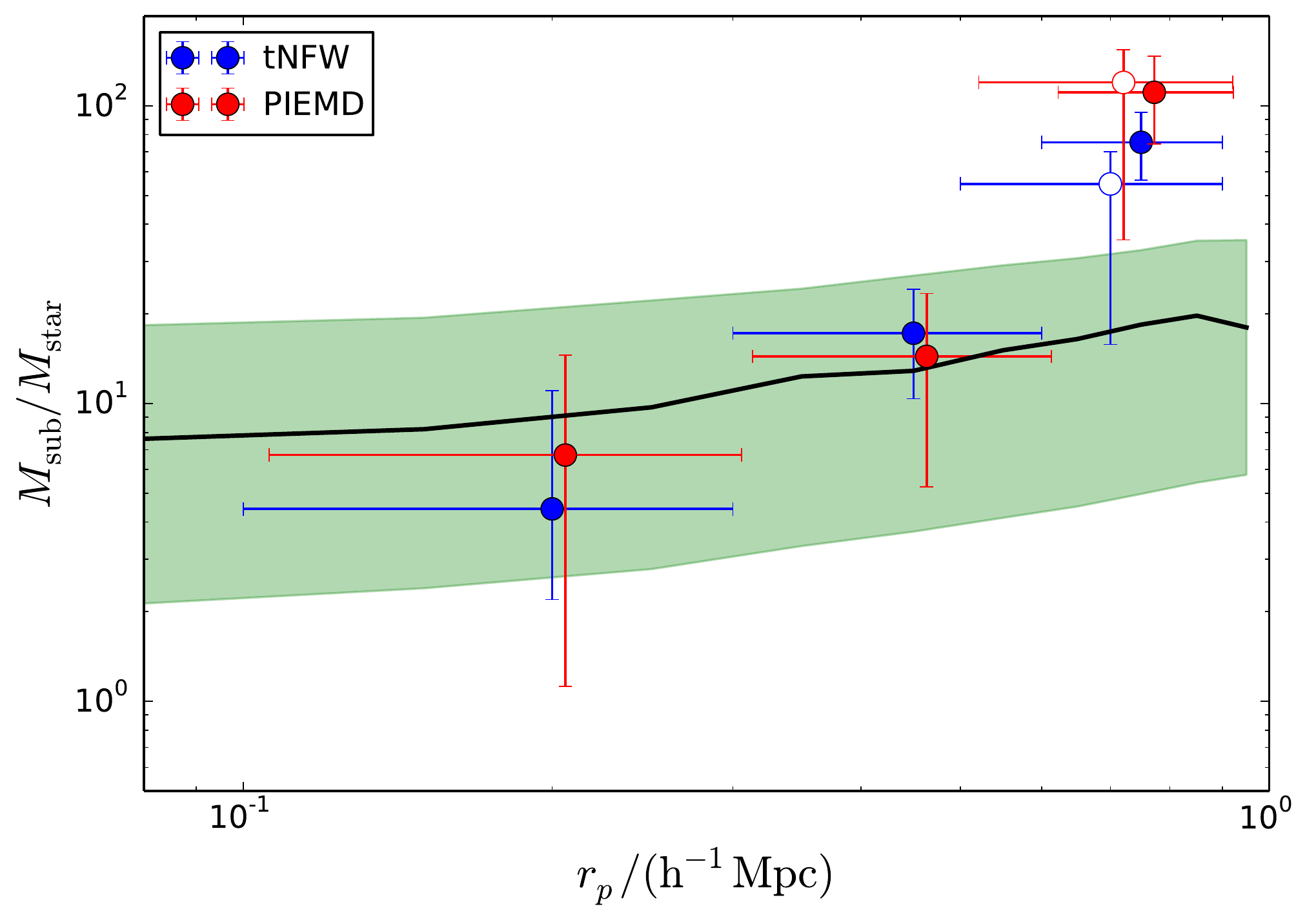}
\caption{The subhalo mass-to-stellar mass ratio for galaxies in different $r_p$ bins. Vertical
error bars show the 68\% confidence interval. Empty circles show the
best-fitted $M_{\rm sub}/M_{\rm star}$ values for $[0.6, 0.9]\mpch$ $r_p$ bin including
the inner most observational  point (see the bottom panel in Fig.\ref{fig:gglensing}).
 Horizontal error bars show the bin range of $r_p$. The green shaded region represents the parameter space where 68\% of the
semi-analytical satellite galaxies distributes. The semi-analytical model also predicts an increasing $M_{\rm sub}/M_{\rm star}$ with $r_p$, but with
a flatter slope.}
\label{fig:ML}
\end{figure}

In numerical simulations, subhalos that are close to host halo centers
are subject to strong mass stripping\citep{Springel2001, DeLucia2004, gao2004, xie2015}.  The mass loss
fraction of subhalos increases from $\sim 30\%$ at $r_{\rm 200}$ to $~90\%$
at 0.1$r_{\rm 200}$ \citep{gao2004,xie2015}. From galaxy-galaxy lensing in 
this work, we also find that the $M_{\rm sub}$ of $[0.6, 0.9]$ $r_p$ bin is
much larger than that in $[0.1, 0.3]$ $r_p$ bin (by a factor of 18).  In
Fig.\ref{fig:ML}, we plot the subhalo mass-to-stellar mass ratio 
for three $r_p$ bins. The $M_{\rm sub}/M_{\rm star}$ ratio
in the $[0.6, 0.9]\mpch$ $r_p$ bin is about 12 times larger than that of the $[0.1, 0.3]\mpch$ bin.
If we include the first point in the $[0.6, 0.9]\mpch$ bin, the $M_{\rm sub}/M_{\rm star}$
of the tNFW model decreases by 40\%.

For reference, we over-plot the $M_{\rm sub}/M_{\rm star}$--$r_p$ predicted by the 
semi-analytical model of \citet{Guo2011}. We adopt the mock galaxy
catalog generated with the \citet{Guo2011} model using the Millennium
simulation \citep{Springel2006}. We select mock satellite galaxies with
stellar masses $M_{\rm star}>10^{10}\ms$ from clusters with $M_{\rm 200}>10^{14}/\ms$.
The median $M_{\rm sub}/M_{\rm star}$-$r_p$ relation is shown in Fig.\ref{fig:ML} with a
black solid line. The green shaded region represents the parameter space where 68\%
of mock satellites distributes. We see the semi-analytical
model predicts an increasing $M_{\rm sub}/M_{\rm star}$ with $r_p$, but with
a flatter slope than in our observations. 
Note that we have not attempted to recreate our detailed observational procedure here,
so source and cluster selection might potentially explain this discrepancy.  
Particularly relevant here is the fact that our analysis relies on a probability cut
$P_{\rm mem}>0.8$ for satellite galaxies, which implies that $\sim 10\%$ of our
satellite galaxies may not be true members of the clusters, but galaxies 
on the line of sight. This contamination is difficult to eliminate completely in 
galaxy-galaxy lensing, because the uncertainties in the line-of-sight distances 
are usually larger than the sizes of the clusters themselves. In \citet{Li2013}, we 
used mock catalogs constructed from $N$-body  simulations to investigate 
the impact of interlopers. It is found that 10\% of the galaxies identified as satellites   
are interlopers, and this introduces a contamination of $15\%$ in the lensing signal.
We expect a comparable level of  bias here, as shown in Appendix A.  
It should be noted, however, that the average membership probability of 
our satellite sample does not  change significantly with $r_p$, implying that 
the contamination by fake member galaxies is similar at different $r_p$. 
We therefore expect that the contamination by interlopers will not lead to  
any qualitative changes in the shapes of the density profiles.   
A detailed comparison between the observation and simulation, 
taking into account the impact of interlopers,  will be carried out in 
a forthcoming paper.

%
\begin{table*}
\begin{center}
  \caption{The best-fit values of the  tNFW model parameters for the stacked
  satellite lensing signal in different $r_p$ bins.
   $\log{M_{\rm sub}} $, and $A_{\rm 0}$ are the best-fit values for the subhalo mass,
    and the normalization factor. We do not
    show the best-fit of $r_s$ and $r_t$ as they are poorly constrained.
     $N_{\rm sat}$  is the number of satellite galaxies in each bin.  $\langle \log M_{\rm star}\rangle$ is the average
      stellar mass of satellites. All errors indicate the 68\% confidence intervals. Masses are in units of $\ms$.
      In our fiducial fitting process, we exclude the first point in $[0.6,0.9]\mpch$ $r_p$ bin as a outlier (see Fig.\ref{fig:gglensing}). 
        For comparison, the bottom row of the table shows the fitting results including this first point. }

\begin{tabular}{l|c|c|c|c|c|c}
 \hline   
 &&&&\\
  $r_p$ range
& $\langle \log(M_{\rm star}) \rangle$ 
& $\log{M_{\rm sub}} $ 
& $A_{\rm 0}$ 
& $M_{\rm sub}/ M_{\rm star}$
&$N_{\rm sat}$ & $\langle z_l \rangle$\\
   &&&&\\
  \hline 
   & & && \\
   $[0.1, 0.3]$ &  10.68 & $ 11.37 ^{+ 0.35 }_{- 0.35}$& $ 0.80^{+ 0.01}_{- 0.01}$ & $ 4.43^{+ 6.63}_{- 2.23}$ & 3963 &0.33 \\
 &  & && \\
 \hline
  &&&&\\
$[0.3, 0.6]$ &  10.72 & $ 11.92 ^{+ 0.19 }_{- 0.18}$& $ 0.86^{+ 0.02}_{- 0.02}$ & $ 17.23^{+ 6.98}_{- 6.84} $ &2507 & 0.29\\
 &  & & &\\
 \hline
  &&&&\\
$[0.6, 0.9]$ &  10.78 & $ 12.64 ^{+ 0.12 }_{- 0.11}$& $ 0.81^{+ 0.04}_{- 0.04}$ & $ 75.40^{+ 19.73}_{- 19.09} $&577 &0.24\\
 &  & & &\\
  \hline
  &&&&\\
$[0.6, 0.9]^*$ &  10.78 & $ 12.49 ^{+ 0.13 }_{- 0.13}$& $ 0.81^{+ 0.04}_{- 0.04}$ & $ 54.64^{+ 15.58}_{- 15.80} $&577 &0.24\\
  &&&&\\
 \hline
 
\end{tabular}
\label{tab:para_nfw}
\end{center}
\end{table*}


\begin{table}
\begin{center}
  \caption{best-fitted parameters with PIEMD model.  }

\begin{tabular}{l|c|c|c|c}
 \hline   
 &&&\\
  $r_p$ range
& $\langle \log(M_{\rm star}) \rangle$ 
& $\log{M_{\rm sub}} $ 
& $A_{\rm 0}$ 
& $M_{\rm sub} /M_{\rm star}$ \\
   &&&\\
  \hline 
   & & & \\ 

$[0.1, 0.3]$ &  10.68 & $ 11.30 ^{+ 0.55 }_{- 0.57}$& $ 0.80^{+ 0.01}_{- 0.01}$ & $ 6.72^{+ 7.84}_{- 5.59} $\\
 &  & & \\
 \hline
  &&&\\
$[0.3, 0.6]$ &  10.72 & $ 11.76 ^{+ 0.33 }_{- 0.32}$& $ 0.86^{+ 0.01}_{- 0.01}$ & $ 14.40^{+ 9.01}_{- 9.16} $\\
 &  & & \\
 \hline
  &&&\\
$[0.6, 0.9]$ &  10.78 & $ 12.80 ^{+ 0.15 }_{- 0.15}$& $ 0.79^{+ 0.04}_{- 0.04}$ & $ 110.98^{+ 35.76}_{- 36.49} $\\
  &&&\\
 \hline
   &&&\\
$[0.6, 0.9]^*$ &  10.78 & $ 12.84 ^{+ 0.13 }_{- 0.13}$& $ 0.79^{+ 0.04}_{- 0.04}$ & $ 119.74^{+ 34.34}_{- 35.47} $\\
   &&&\\
 \hline
\end{tabular}
\label{tab:para_piemd}
\end{center}
\end{table}


\subsection{Dependence on satellite stellar mass}

In the CDM structure formation scenario, satellite galaxies with larger stellar mass tend to occupy larger
haloes at infall time \citep[e.g.][]{Vale2004, Conroy2006,Yang_etal2012,Lu_etal2014}.
In addition, massive haloes may retain more mass than lower mass ones 
at the same halo-centric radius \citep[e.g.][]{Conroy2006,Moster2010,Simha2012}. 
For these two reasons,  we expect that satellite galaxies with larger stellar 
mass reside in more massive subhalos.

\begin{figure}
\includegraphics[width=0.5\textwidth]{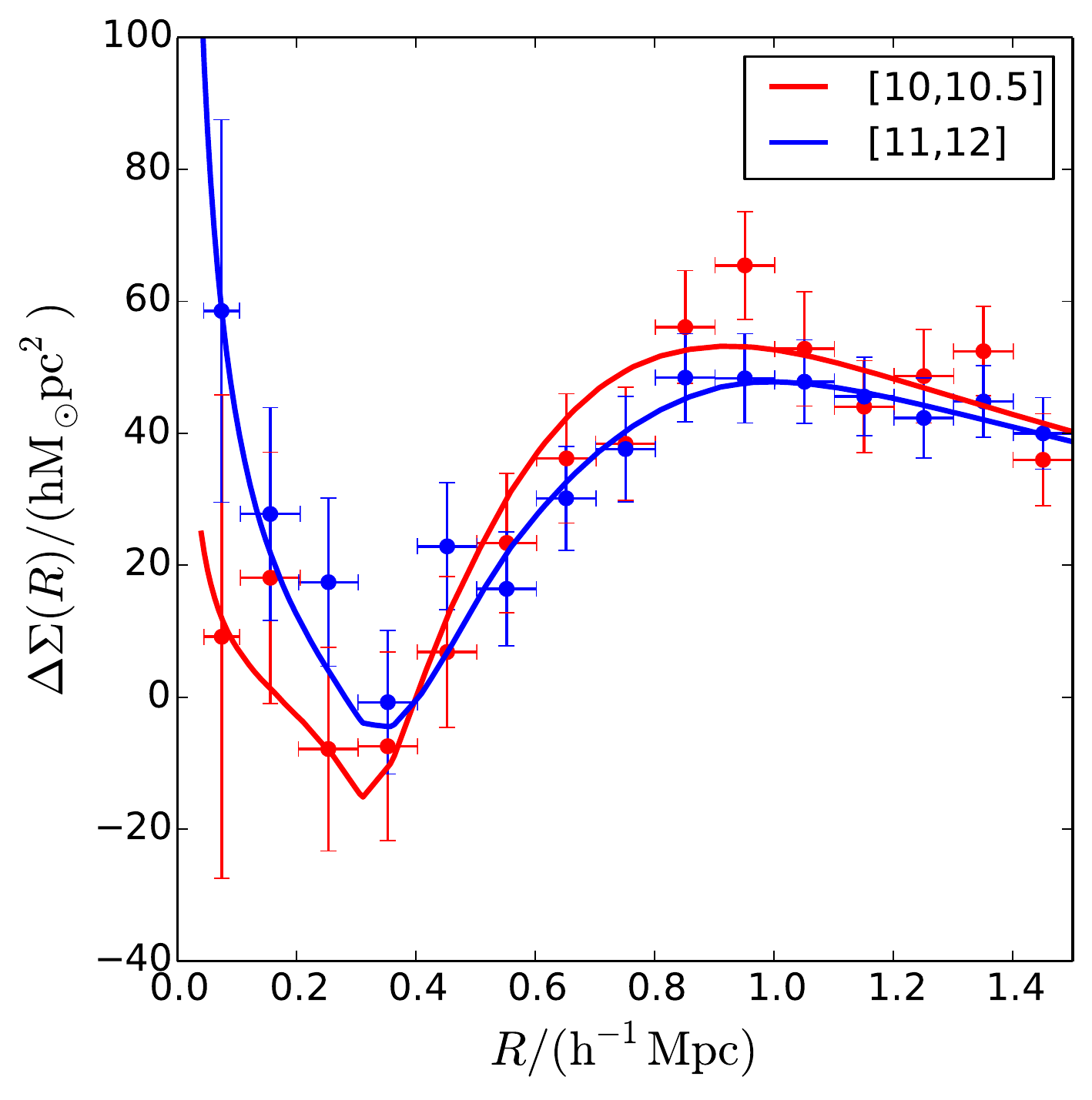}
\caption{Observed galaxy-galaxy lensing signal  for  satellite
  galaxies in different stellar mass bins at fixed $r_p$. The
legend shows the $\log(M_{\rm star}/\ms)$ range for different data points. 
The solid lines are the best-fit tNFW models.}
\label{fig:mstar}
\end{figure}

To test this prediction, we select all galaxies with $r_p$ in the range $[0.3, 0.9]$ $\mpch$,  and split
the them into two subsamples: $10<\log(M_{\rm star}/\ms)<10.5$ and $11<\log(M_{\rm star}/\ms)<12$.
The galaxy-galaxy lensing signals of the two satellite samples are shown in Fig.\ref{fig:mstar}. It is clear
that at small scales, where subhalos dominate, the lensing signals are larger around the more massive
satellites. The best-fit subhalo  mass for the high mass and low mass satellites are
$\log(M_{\rm sub}/\ms)=11.14 ^{+ 0.66 }_{- 0.73}$ ($M_{\rm sub}/M_{\rm star}=19.5^{+19.8}_{-17.9}$) and
$\log(M_{\rm sub}/\ms)=12.38 ^{+ 0.16 }_{- 0.16}$ ($M_{\rm sub}/M_{\rm star}=21.1^{+7.4}_{-7.7}$ respectively.

\section{Discussion and conclusion}
\label{sec:sum}

In this paper, we present measurements of the galaxy-galaxy lensing signal of satellite galaxies in redMaPPer
clusters. We select satellite galaxies from massive clusters (richness $\lambda>20$) in the redMaPPer catalog. 
We fit the galaxy-galaxy lensing signal around the satellites using a tNFW profile and a PIEMD profile, and obtain the  subhalo masses. 
We bin satellite galaxies according to their projected halo-centric distance $r_p$
and find that the best-fit subhalo mass of satellite galaxies increases with $r_p$. The best-fit $\log({M_{\rm sub}/\ms})$ 
for satellites in the $r_p=[0.6, 0.9]\mpch$ bin is larger than in the $r_p=[0.1, 0.3]\mpch$ bin by a factor of 18.
The $M_{\rm sub}/M_{\rm star}$ ratio is also found to increase as a function of $r_p$ by a factor of 12. We also find that satellites with more stellar
mass tend to populate more massive subhalos. Our results provide evidence for the tidal striping effects on the subhalos of red sequence satellite galaxies,
as expected on the basis of the CDM hierarchical structure formation scenario. 

Many previous studies have tried to test this theoretical prediction
using gravitational lensing. Most of these previous studies focus on measuring
subhalo mass in individual rich clusters. For example, \citet{Natarajan2009} report
the measurement of dark matter subhalos as a function of projected
halo-centric radius for the cluster C10024+16. They found that the mass of
dark matter subhalos hosting early type $L^{*}$ galaxies increases
by a factor of 6 from a halo-centric radius $r<0.6$ $\mpch$ to $r\sim 4$
$\mpch$.  In recent work, \citet{Okabe2014} present the weak
lensing measurement of 32 subhalos in the very nearby Coma cluster.  They
also found that the mass-to-light ratio of subhalos increase as a
function of halo-centric radius: the $M/L_{\rm i'}$ of subhalos
increases from 60 at 10$'$ to about 500 at 70$'$. 

Our work is complementary to these results. In the paper, 
we measure the galaxy-galaxy lensing signal of subhalos in a statistical sample of rich
clusters. Our results lead to similar conclusions and show evidence
for the effects of tidal striping.  However, due to the statistical noise of the current survey,
there is still a large uncertainties in the measurement of $M_{\rm
  sub}/M_{\rm star}$.
  
Theoretically, the galaxy - halo connection can be established 
by studying how galaxies of different properties reside in dark matter 
halos of different masses, through models  such as the halo occupation 
models \citet[e.g.][]{Jing_etal1998, PeacockSmith2000, 
BerlindWeinberg2002}, the conditional luminosity functions models
\citep[e.g.][]{Yang_etal2003}, (sub)halo-abundance-matching 
\citep[ e.g.][]{Vale2004, Conroy2006, Wang2006, Yang_etal2012, 
Chaves-Montero2015}, and empirical models of star formation and mass assembly 
of galaxies in dark matter halos \citep[e.g.][]{Lu_etal2014, Lu_etal2015}. 
In most of these models, the connection is made between galaxy luminosities
(stellar masses) and halo masses before a galaxy becomes a satellite, i.e. 
before a galaxy and its halo have experienced halo-specific 
environmental effects. The tidal stripping effects after a halo becomes 
a sub-halo can be followed in dark matter simulations. 
This, together with the halo-galaxy connection established through 
the various models,  can be used to predict the subhalo - galaxy 
relation at the present day. During the revision of this paper, 
\citet{Han2016} posted a theoretical work on subhalo spatial and mass 
distributions. In Sec. 6.4 of their paper, they presented how our lensing 
measurements can be derived theoretically from a subhalo abundance 
matching point of view.  With future data, the galaxy-galaxy lensing 
measurements of subhalos associated with satellites are expected 
to provide important constraints on galaxy formation and evolution in 
dark matter halos.

The results of this paper also demonstrate the promise of next generation weak 
lensing surveys. In \citet{Li2013}, it is shown that the constraints on subhalo 
structures and $M_{\rm sub}/M_{\rm star}$ can be improved dramatically with 
next generation galaxy surveys such as LSST, due to the increase in both sky 
area (17000 deg$^2$) and the depth (40 gal/arcmin$^2$), which is crucial 
in constraining the co-evolution of satellite galaxies and subhalos. 
The space mission, such as Euclid, will survey a similar area of the 
sky (20000 deg$^2$) but with much better image resolutions 
(FWHM $0.1"$ versus $0.7"$  for LSST), which is expected to 
provide even better measurements of galaxy-galaxy lensing. The method 
described  here can readily be extended to these future surveys.

\section*{Acknowledgements}

Based on observations obtained with MegaPrime/MegaCam, a joint project of CFHT and CEA/DAPNIA, 
at the Canada-France-Hawaii Telescope (CFHT), which is operated by the National Research 
Council (NRC) of Canada, the Institut National des Science de l'Univers of the Centre National de la 
Recherche Scientifique (CNRS) of France, and the University of Hawaii. The Brazilian 
partnership on CFHT is managed by the Laborat\'orio Nacional de Astrof\`isica (LNA). 
This work made use of the CHE cluster, managed and funded by ICRA/CBPF/MCTI, 
with financial support from FINEP and FAPERJ. We thank the support of the 
Laboratrio Interinstitucional de e-Astronomia (LIneA). We thank the CFHTLenS 
team for their pipeline development and verification upon which much of this surveys pipeline was built.

LR acknowledges the NSFC(grant No.11303033,11133003), the support from 
Youth Innovation Promotion Association of CAS. HYS acknowledges the support from 
Marie-Curie International Fellowship (FP7-PEOPLE-2012-IIF/327561), Swiss National 
Science Foundation (SNSF) and NSFC of China under grants 11103011.
HJM acknowledges support of NSF AST-0908334, NSF AST-1109354 and NSF AST-1517528.
JPK acknowledges support from the ERC advanced grant LIDA and from CNRS.
TE is supported by the Deutsche Forschungs-gemeinschaft through the 
Transregional Collaborative Research Centre TR 33 - The Dark Universe.
AL is supported by World Premier International Research Center Initiative (WPI Initiative), MEXT, Japan.
BM acknowledges financial support from the CAPES Foundation grant 12174-13-0.
MM is partially supported by CNPq (grant 486586/2013-8) and FAPERJ (grant E-26/110.516/2-2012).

\bibliography{lensing}

\begin{thebibliography}{}

\bibitem[\protect\citeauthoryear{{Abadi}, {Moore} \& {Bower}}{{Abadi}
  et~al.}{1999}]{Abadi1999}
{Abadi} M.~G.,  {Moore} B.,    {Bower} R.~G.,  1999, \mnras, 308, 947

\bibitem[\protect\citeauthoryear{{Aihara}, {Allende Prieto}, {An}, {Anderson},
  {Aubourg}, {Balbinot} \& {Beers}}{{Aihara} et~al.}{2011}]{Aihara2011}
{Aihara} H.,  {Allende Prieto} C.,  {An} D.,  {Anderson} S.~F.,  {Aubourg}
  {\'E}.,  {Balbinot} E.,    {Beers} T.~C.,  2011, \apjs, 195, 26

\bibitem[\protect\citeauthoryear{{Annis}, {Soares-Santos}, {Strauss}, {Becker},
  {Dodelson}, {Fan}, {Gunn}, {Hao}, {Ivezi{\'c}}, {Jester}, {Jiang},
  {Johnston}, {Kubo}, {Lampeitl}, {Lin}, {Lupton}, {Miknaitis}, {Seo}, {Simet}
  \& {Yanny}}{{Annis} et~al.}{2014}]{Annis2014}
{Annis} J.,  {Soares-Santos} M.,  {Strauss} M.~A.,  {Becker} A.~C.,  {Dodelson}
  S.,  {Fan} X.,  {Gunn} J.~E.,  {Hao} J.,  {Ivezi{\'c}} {\v Z}.,  {Jester} S.,
   {Jiang} L.,  {Johnston} D.~E.,  {Kubo} J.~M.,  {Lampeitl} H.,  {Lin} H.,
  {Lupton} R.~H.,  {Miknaitis} G.,  {Seo} H.-J.,  {Simet} M.,    {Yanny} B.,
  2014, \apj, 794, 120

\bibitem[\protect\citeauthoryear{{Balogh}, {Navarro} \& {Morris}}{{Balogh}
  et~al.}{2000}]{Balogh2000}
{Balogh} M.~L.,  {Navarro} J.~F.,    {Morris} S.~L.,  2000, \apj, 540, 113

\bibitem[\protect\citeauthoryear{{Baltz}, {Marshall} \& {Oguri}}{{Baltz}
  et~al.}{2009}]{Baltz2009}
{Baltz} E.~A.,  {Marshall} P.,    {Oguri} M.,  2009, \jcap, 1, 15

\bibitem[\protect\citeauthoryear{{Ben{\'{\i}}tez}}{{Ben{\'{\i}}tez}}{2000}]{Benitez2000}
{Ben{\'{\i}}tez} N.,  2000, \apj, 536, 571

\bibitem[\protect\citeauthoryear{{Berlind} \& {Weinberg}}{{Berlind} \&
  {Weinberg}}{2002}]{BerlindWeinberg2002}
{Berlind} A.~A.,  {Weinberg} D.~H.,  2002, \apj, 575, 587

\bibitem[\protect\citeauthoryear{{Bose}, {Hellwing}, {Frenk}, {Jenkins},
  {Lovell}, {Helly} \& {Li}}{{Bose} et~al.}{2016}]{Bose2016}
{Bose} S.,  {Hellwing} W.~A.,  {Frenk} C.~S.,  {Jenkins} A.,  {Lovell} M.~R.,
  {Helly} J.~C.,    {Li} B.,  2016, \mnras, 455, 318

\bibitem[\protect\citeauthoryear{{Bullock}, {Kolatt}, {Sigad}, {Somerville},
  {Kravtsov}, {Klypin}, {Primack} \& {Dekel}}{{Bullock}
  et~al.}{2001}]{Bullock2001}
{Bullock} J.~S.,  {Kolatt} T.~S.,  {Sigad} Y.,  {Somerville} R.~S.,  {Kravtsov}
  A.~V.,  {Klypin} A.~A.,  {Primack} J.~R.,    {Dekel} A.,  2001, \mnras, 321,
  559

\bibitem[\protect\citeauthoryear{{Bundy}, {Leauthaud}, {Saito}, {Bolton},
  {Lin}, {Marason}, {Nichol}, {Schneider}, {Thomas} \& {Wake}}{{Bundy}
  et~al.}{2015}]{Bundy2015}
{Bundy} K.,  {Leauthaud} A.,  {Saito} S.,  {Bolton} A.,  {Lin} Y.-T.,
  {Marason} C.,  {Nichol} R.~C.,  {Schneider} D.~P.,  {Thomas} D.,    {Wake}
  D.~A.,  2015, ArXiv e-prints

\bibitem[\protect\citeauthoryear{{Chabrier}}{{Chabrier}}{2003}]{Chabrier2003}
{Chabrier} G.,  2003, \pasp, 115, 763

\bibitem[\protect\citeauthoryear{{Chang}, {Macci{\`o}} \& {Kang}}{{Chang}
  et~al.}{2013}]{Chang2013}
{Chang} J.,  {Macci{\`o}} A.~V.,    {Kang} X.,  2013, \mnras, 431, 3533

\bibitem[\protect\citeauthoryear{{Chaves-Montero}, {Angulo}, {Schaye},
  {Schaller}, {Crain} \& {Furlong}}{{Chaves-Montero}
  et~al.}{2015}]{Chaves-Montero2015}
{Chaves-Montero} J.,  {Angulo} R.~E.,  {Schaye} J.,  {Schaller} M.,  {Crain}
  R.~A.,    {Furlong} M.,  2015, ArXiv e-prints

\bibitem[\protect\citeauthoryear{{Chung}, {van Gorkom}, {Kenney}, {Crowl} \&
  {Vollmer}}{{Chung} et~al.}{2009}]{Chung2009}
{Chung} A.,  {van Gorkom} J.~H.,  {Kenney} J.~D.~P.,  {Crowl} H.,    {Vollmer}
  B.,  2009, \aj, 138, 1741

\bibitem[\protect\citeauthoryear{{Col{\'{\i}}n}, {Avila-Reese} \&
  {Valenzuela}}{{Col{\'{\i}}n} et~al.}{2000}]{Colin2000}
{Col{\'{\i}}n} P.,  {Avila-Reese} V.,    {Valenzuela} O.,  2000, \apj, 542, 622

\bibitem[\protect\citeauthoryear{{Comparat}, {Jullo}, {Kneib}, {Schimd},
  {Shan}, {Erben}, {Ilbert}, {Brownstein}, {Ealet}, {Escoffier}, {Moraes},
  {Mostek}, {Newman}, {Pereira}, {Prada}, {Schlegel}, {Schneider} \&
  {Brandt}}{{Comparat} et~al.}{2013}]{Comparat2013}
{Comparat} J.,  {Jullo} E.,  {Kneib} J.-P.,  {Schimd} C.,  {Shan} H.,  {Erben}
  T.,  {Ilbert} O.,  {Brownstein} J.,  {Ealet} A.,  {Escoffier} S.,  {Moraes}
  B.,  {Mostek} N.,  {Newman} J.~A.,  {Pereira} M.~E.~S.,  {Prada} F.,
  {Schlegel} D.~J.,  {Schneider} D.~P.,    {Brandt} C.~H.,  2013, \mnras, 433,
  1146

\bibitem[\protect\citeauthoryear{{Conroy}, {Gunn} \& {White}}{{Conroy}
  et~al.}{2009}]{Conroy2009}
{Conroy} C.,  {Gunn} J.~E.,    {White} M.,  2009, \apj, 699, 486

\bibitem[\protect\citeauthoryear{{Conroy}, {Wechsler} \& {Kravtsov}}{{Conroy}
  et~al.}{2006}]{Conroy2006}
{Conroy} C.,  {Wechsler} R.~H.,    {Kravtsov} A.~V.,  2006, \apj, 647, 201

\bibitem[\protect\citeauthoryear{{Conroy}, {White} \& {Gunn}}{{Conroy}
  et~al.}{2010}]{Conroy2010}
{Conroy} C.,  {White} M.,    {Gunn} J.~E.,  2010, \apj, 708, 58

\bibitem[\protect\citeauthoryear{{Dawson}, {Schlegel}, {Ahn}, {Anderson},
  {Aubourg}, {Bailey}, {Barkhouser} \& {Bautista}}{{Dawson}
  et~al.}{2013}]{Dawson2013}
{Dawson} K.~S.,  {Schlegel} D.~J.,  {Ahn} C.~P.,  {Anderson} S.~F.,  {Aubourg}
  {\'E}.,  {Bailey} S.,  {Barkhouser} R.~H.,    {Bautista} J.~E.,  2013, \aj,
  145, 10

\bibitem[\protect\citeauthoryear{{De Lucia}, {Kauffmann}, {Springel}, {White},
  {Lanzoni}, {Stoehr}, {Tormen} \& {Yoshida}}{{De Lucia}
  et~al.}{2004}]{DeLucia2004}
{De Lucia} G.,  {Kauffmann} G.,  {Springel} V.,  {White} S.~D.~M.,  {Lanzoni}
  B.,  {Stoehr} F.,  {Tormen} G.,    {Yoshida} N.,  2004, \mnras, 348, 333

\bibitem[\protect\citeauthoryear{{Diemand}, {Kuhlen} \& {Madau}}{{Diemand}
  et~al.}{2007}]{Diemand2007}
{Diemand} J.,  {Kuhlen} M.,    {Madau} P.,  2007, \apj, 657, 262

\bibitem[\protect\citeauthoryear{{Dolag}, {Bartelmann}, {Perrotta},
  {Baccigalupi}, {Moscardini}, {Meneghetti} \& {Tormen}}{{Dolag}
  et~al.}{2004}]{Dolag2004}
{Dolag} K.,  {Bartelmann} M.,  {Perrotta} F.,  {Baccigalupi} C.,  {Moscardini}
  L.,  {Meneghetti} M.,    {Tormen} G.,  2004, \aap, 416, 853

\bibitem[\protect\citeauthoryear{{Duffy}, {Schaye}, {Kay} \& {Dalla
  Vecchia}}{{Duffy} et~al.}{2008}]{Duffy2008}
{Duffy} A.~R.,  {Schaye} J.,  {Kay} S.~T.,    {Dalla Vecchia} C.,  2008,
  \mnras, 390, L64

\bibitem[\protect\citeauthoryear{{Erben}, {Hildebrandt}, {Lerchster},
  {Hudelot}, {Benjamin}, {van Waerbeke}, {Schrabback} \& {Brimioulle}}{{Erben}
  et~al.}{2009}]{Erben2009}
{Erben} T.,  {Hildebrandt} H.,  {Lerchster} M.,  {Hudelot} P.,  {Benjamin} J.,
  {van Waerbeke} L.,  {Schrabback} T.,    {Brimioulle} F.,  2009, \aap, 493,
  1197

\bibitem[\protect\citeauthoryear{{Erben}, {Hildebrandt}, {Miller}, {van
  Waerbeke}, {Heymans}, {Hoekstra}, {Kitching} \& {Mellier}}{{Erben}
  et~al.}{2013}]{Erben2013}
{Erben} T.,  {Hildebrandt} H.,  {Miller} L.,  {van Waerbeke} L.,  {Heymans} C.,
   {Hoekstra} H.,  {Kitching} T.~D.,    {Mellier} Y.,  2013, \mnras, 433, 2545

\bibitem[\protect\citeauthoryear{{Frenk} \& {White}}{{Frenk} \&
  {White}}{2012}]{Frenk2012}
{Frenk} C.~S.,  {White} S.~D.~M.,  2012, Annalen der Physik, 524, 507

\bibitem[\protect\citeauthoryear{{Gao}, {Frenk}, {Boylan-Kolchin}, {Jenkins},
  {Springel} \& {White}}{{Gao} et~al.}{2011}]{Gao2011}
{Gao} L.,  {Frenk} C.~S.,  {Boylan-Kolchin} M.,  {Jenkins} A.,  {Springel} V.,
    {White} S.~D.~M.,  2011, \mnras, 410, 2309

\bibitem[\protect\citeauthoryear{{Gao}, {Navarro}, {Frenk}, {Jenkins},
  {Springel} \& {White}}{{Gao} et~al.}{2012}]{Gao2012}
{Gao} L.,  {Navarro} J.~F.,  {Frenk} C.~S.,  {Jenkins} A.,  {Springel} V.,
  {White} S.~D.~M.,  2012, \mnras, 425, 2169

\bibitem[\protect\citeauthoryear{{Gao}, {White}, {Jenkins}, {Stoehr} \&
  {Springel}}{{Gao} et~al.}{2004}]{gao2004}
{Gao} L.,  {White} S.~D.~M.,  {Jenkins} A.,  {Stoehr} F.,    {Springel} V.,
  2004, \mnras, 355, 819

\bibitem[\protect\citeauthoryear{{George}, {Leauthaud}, {Bundy}, {Finoguenov},
  {Ma}, {Rykoff}, {Tinker} \& {Wechsler}}{{George} et~al.}{2012}]{George2012}
{George} M.~R.,  {Leauthaud} A.,  {Bundy} K.,  {Finoguenov} A.,  {Ma} C.-P.,
  {Rykoff} E.~S.,  {Tinker} J.~L.,    {Wechsler} R.~H.,  2012, \apj, 757, 2

\bibitem[\protect\citeauthoryear{{Gillis}, {Hudson}, {Erben}, {Heymans},
  {Hildebrandt}, {Hoekstra}, {Kitching} \& {Mellier}}{{Gillis}
  et~al.}{2013}]{Gillis2013b}
{Gillis} B.~R.,  {Hudson} M.~J.,  {Erben} T.,  {Heymans} C.,  {Hildebrandt} H.,
   {Hoekstra} H.,  {Kitching} T.~D.,    {Mellier} 2013, \mnras, 431, 1439

\bibitem[\protect\citeauthoryear{{Gunn} \& {Gott} III}{{Gunn} \&
  {Gott}}{1972}]{Gunn1972}
{Gunn} J.~E.,  {Gott} III J.~R.,  1972, \apj, 176, 1

\bibitem[\protect\citeauthoryear{{Guo}, {White}, {Boylan-Kolchin}, {De Lucia},
  {Kauffmann}, {Lemson}, {Li}, {Springel} \& {Weinmann}}{{Guo}
  et~al.}{2011}]{Guo2011}
{Guo} Q.,  {White} S.,  {Boylan-Kolchin} M.,  {De Lucia} G.,  {Kauffmann} G.,
  {Lemson} G.,  {Li} C.,  {Springel} V.,    {Weinmann} S.,  2011, \mnras, 413,
  101

\bibitem[\protect\citeauthoryear{{Han}, {Cole}, {Frenk} \& {Jing}}{{Han}
  et~al.}{2016}]{Han2016}
{Han} J.,  {Cole} S.,  {Frenk} C.~S.,    {Jing} Y.,  2016, \mnras, 457, 1208

\bibitem[\protect\citeauthoryear{{Han}, {Eke}, {Frenk}, {Mandelbaum},
  {Norberg}, {Schneider}, {Peacock}, {Jing}, {Baldry}, {Bland-Hawthorn},
  {Brough}, {Brown}, {Liske}, {Loveday} \& {Robotham}}{{Han}
  et~al.}{2015}]{Han2015}
{Han} J.,  {Eke} V.~R.,  {Frenk} C.~S.,  {Mandelbaum} R.,  {Norberg} P.,
  {Schneider} M.~D.,  {Peacock} J.~A.,  {Jing} Y.,  {Baldry} I.,
  {Bland-Hawthorn} J.,  {Brough} S.,  {Brown} M.~J.~I.,  {Liske} J.,  {Loveday}
  J.,    {Robotham} A.~S.~G.,  2015, \mnras, 446, 1356

\bibitem[\protect\citeauthoryear{{Hayashi}, {Navarro}, {Taylor}, {Stadel} \&
  {Quinn}}{{Hayashi} et~al.}{2003}]{Hayashi2003}
{Hayashi} E.,  {Navarro} J.~F.,  {Taylor} J.~E.,  {Stadel} J.,    {Quinn} T.,
  2003, \apj, 584, 541

\bibitem[\protect\citeauthoryear{{Hellwing}, {Frenk}, {Cautun}, {Bose},
  {Helly}, {Jenkins}, {Sawala} \& {Cytowski}}{{Hellwing}
  et~al.}{2015}]{Hellwing2015}
{Hellwing} W.~A.,  {Frenk} C.~S.,  {Cautun} M.,  {Bose} S.,  {Helly} J.,
  {Jenkins} A.,  {Sawala} T.,    {Cytowski} M.,  2015, ArXiv e-prints

\bibitem[\protect\citeauthoryear{{Heymans}, {Van Waerbeke}, {Miller}, {Erben},
  {Hildebrandt}, {Hoekstra}, {Kitching} \& {Mellier}}{{Heymans}
  et~al.}{2012}]{Heymans2012}
{Heymans} C.,  {Van Waerbeke} L.,  {Miller} L.,  {Erben} T.,  {Hildebrandt} H.,
   {Hoekstra} H.,  {Kitching} T.~D.,    {Mellier} Y.,  2012, \mnras, 427, 146

\bibitem[\protect\citeauthoryear{{Jing}, {Mo} \& {B{\"o}rner}}{{Jing}
  et~al.}{1998}]{Jing_etal1998}
{Jing} Y.~P.,  {Mo} H.~J.,    {B{\"o}rner} G.,  1998, \apj, 494, 1

\bibitem[\protect\citeauthoryear{{Kang} \& {van den Bosch}}{{Kang} \& {van den
  Bosch}}{2008}]{Kang2008}
{Kang} X.,  {van den Bosch} F.~C.,  2008, \apjl, 676, L101

\bibitem[\protect\citeauthoryear{{Kassiola} \& {Kovner}}{{Kassiola} \&
  {Kovner}}{1993}]{Kassiola1993}
{Kassiola} A.,  {Kovner} I.,  1993, \apj, 417, 450

\bibitem[\protect\citeauthoryear{{Kawata} \& {Mulchaey}}{{Kawata} \&
  {Mulchaey}}{2008}]{Kawata2008}
{Kawata} D.,  {Mulchaey} J.~S.,  2008, \apjl, 672, L103

\bibitem[\protect\citeauthoryear{{Klimentowski}, {{\L}okas}, {Kazantzidis},
  {Prada}, {Mayer} \& {Mamon}}{{Klimentowski} et~al.}{2007}]{Klimentowski2007}
{Klimentowski} J.,  {{\L}okas} E.~L.,  {Kazantzidis} S.,  {Prada} F.,  {Mayer}
  L.,    {Mamon} G.~A.,  2007, \mnras, 378, 353

\bibitem[\protect\citeauthoryear{{Kneib}, {Ellis}, {Smail}, {Couch} \&
  {Sharples}}{{Kneib} et~al.}{1996}]{kneib1996}
{Kneib} J.-P.,  {Ellis} R.~S.,  {Smail} I.,  {Couch} W.~J.,    {Sharples}
  R.~M.,  1996, \apj, 471, 643

\bibitem[\protect\citeauthoryear{{Kneib} \& {Natarajan}}{{Kneib} \&
  {Natarajan}}{2011}]{Kneib2011}
{Kneib} J.-P.,  {Natarajan} P.,  2011, \aapr, 19, 47

\bibitem[\protect\citeauthoryear{{Komatsu}, {Smith}, {Dunkley}, {Bennett},
  {Gold}, {Hinshaw}, {Jarosik} \& {Larson}}{{Komatsu}
  et~al.}{2010}]{komatsu2010}
{Komatsu} E.,  {Smith} K.~M.,  {Dunkley} J.,  {Bennett} C.~L.,  {Gold} B.,
  {Hinshaw} G.,  {Jarosik} N.,    {Larson} D.,  2010, ArXiv e-prints

\bibitem[\protect\citeauthoryear{{Koopmans}}{{Koopmans}}{2005}]{Koopmans2005}
{Koopmans} L.~V.~E.,  2005, \mnras, 363, 1136

\bibitem[\protect\citeauthoryear{{Lang}, {Hogg} \& {Schlegel}}{{Lang}
  et~al.}{2014}]{Lang2014}
{Lang} D.,  {Hogg} D.~W.,    {Schlegel} D.~J.,  2014, ArXiv e-prints

\bibitem[\protect\citeauthoryear{{Li}, {Mo}, {Fan}, {Cacciato}, {van den
  Bosch}, {Yang} \& {More}}{{Li} et~al.}{2009}]{Li2009}
{Li} R.,  {Mo} H.~J.,  {Fan} Z.,  {Cacciato} M.,  {van den Bosch} F.~C.,
  {Yang} X.,    {More} S.,  2009, \mnras, 394, 1016

\bibitem[\protect\citeauthoryear{{Li}, {Mo}, {Fan}, {Yang} \& {Bosch}}{{Li}
  et~al.}{2013}]{Li2013}
{Li} R.,  {Mo} H.~J.,  {Fan} Z.,  {Yang} X.,    {Bosch} F.~C.~v.~d.,  2013,
  \mnras, 430, 3359

\bibitem[\protect\citeauthoryear{{Li}, {Shan}, {Mo}, {Kneib}, {Yang}, {Luo},
  {van den Bosch}, {Erben}, {Moraes} \& {Makler}}{{Li} et~al.}{2014}]{Li2014}
{Li} R.,  {Shan} H.,  {Mo} H.,  {Kneib} J.-P.,  {Yang} X.,  {Luo} W.,  {van den
  Bosch} F.~C.,  {Erben} T.,  {Moraes} B.,    {Makler} M.,  2014, \mnras, 438,
  2864

\bibitem[\protect\citeauthoryear{{Limousin}, {Kneib}, {Bardeau}, {Natarajan},
  {Czoske}, {Smail}, {Ebeling} \& {Smith}}{{Limousin}
  et~al.}{2007}]{Limousin2007}
{Limousin} M.,  {Kneib} J.~P.,  {Bardeau} S.,  {Natarajan} P.,  {Czoske} O.,
  {Smail} I.,  {Ebeling} H.,    {Smith} G.~P.,  2007, \aap, 461, 881

\bibitem[\protect\citeauthoryear{{Limousin}, {Kneib} \& {Natarajan}}{{Limousin}
  et~al.}{2005}]{Limousin2005}
{Limousin} M.,  {Kneib} J.-P.,    {Natarajan} P.,  2005, \mnras, 356, 309

\bibitem[\protect\citeauthoryear{{Limousin}, {Sommer-Larsen}, {Natarajan} \&
  {Milvang-Jensen}}{{Limousin} et~al.}{2009}]{Limousin2009}
{Limousin} M.,  {Sommer-Larsen} J.,  {Natarajan} P.,    {Milvang-Jensen} B.,
  2009, \apj, 696, 1771

\bibitem[\protect\citeauthoryear{{Lu}, {Mo}, {Lu}, {Katz}, {Weinberg}, {van den
  Bosch} \& {Yang}}{{Lu} et~al.}{2014}]{Lu_etal2014}
{Lu} Z.,  {Mo} H.~J.,  {Lu} Y.,  {Katz} N.,  {Weinberg} M.~D.,  {van den Bosch}
  F.~C.,    {Yang} X.,  2014, \mnras, 439, 1294

\bibitem[\protect\citeauthoryear{{Lu}, {Mo}, {Lu}, {Katz}, {Weinberg}, {van den
  Bosch} \& {Yang}}{{Lu} et~al.}{2015}]{Lu_etal2015}
{Lu} Z.,  {Mo} H.~J.,  {Lu} Y.,  {Katz} N.,  {Weinberg} M.~D.,  {van den Bosch}
  F.~C.,    {Yang} X.,  2015, \mnras, 450, 1604

\bibitem[\protect\citeauthoryear{{Macci{\`o}}, {Dutton}, {van den Bosch},
  {Moore}, {Potter} \& {Stadel}}{{Macci{\`o}} et~al.}{2007}]{Maccio2007}
{Macci{\`o}} A.~V.,  {Dutton} A.~A.,  {van den Bosch} F.~C.,  {Moore} B.,
  {Potter} D.,    {Stadel} J.,  2007, \mnras, 378, 55

\bibitem[\protect\citeauthoryear{{Mandelbaum}, {Hirata}, {Seljak}, {Guzik},
  {Padmanabhan}, {Blake}, {Blanton}, {Lupton} \& {Brinkmann}}{{Mandelbaum}
  et~al.}{2005}]{Mandelbaum2005}
{Mandelbaum} R.,  {Hirata} C.~M.,  {Seljak} U.,  {Guzik} J.,  {Padmanabhan} N.,
   {Blake} C.,  {Blanton} M.~R.,  {Lupton} R.,    {Brinkmann} J.,  2005,
  \mnras, 361, 1287

\bibitem[\protect\citeauthoryear{{Mandelbaum}, {Seljak} \&
  {Hirata}}{{Mandelbaum} et~al.}{2008}]{Mandelbaum2008}
{Mandelbaum} R.,  {Seljak} U.,    {Hirata} C.~M.,  2008, \jcap, 8, 6

\bibitem[\protect\citeauthoryear{{Mao}, {Jing}, {Ostriker} \& {Weller}}{{Mao}
  et~al.}{2004}]{Mao2004}
{Mao} S.,  {Jing} Y.,  {Ostriker} J.~P.,    {Weller} J.,  2004, \apjl, 604, L5

\bibitem[\protect\citeauthoryear{{Mao} \& {Schneider}}{{Mao} \&
  {Schneider}}{1998}]{Mao1998}
{Mao} S.,  {Schneider} P.,  1998, \mnras, 295, 587

\bibitem[\protect\citeauthoryear{{Mayer}, {Governato}, {Colpi}, {Moore},
  {Quinn}, {Wadsley}, {Stadel} \& {Lake}}{{Mayer} et~al.}{2001}]{Mayer2001}
{Mayer} L.,  {Governato} F.,  {Colpi} M.,  {Moore} B.,  {Quinn} T.,  {Wadsley}
  J.,  {Stadel} J.,    {Lake} G.,  2001, \apjl, 547, L123

\bibitem[\protect\citeauthoryear{{McCarthy}, {Frenk}, {Font}, {Lacey}, {Bower},
  {Mitchell}, {Balogh} \& {Theuns}}{{McCarthy} et~al.}{2008}]{McCarthy2008}
{McCarthy} I.~G.,  {Frenk} C.~S.,  {Font} A.~S.,  {Lacey} C.~G.,  {Bower}
  R.~G.,  {Mitchell} N.~L.,  {Balogh} M.~L.,    {Theuns} T.,  2008, \mnras,
  383, 593

\bibitem[\protect\citeauthoryear{{Metcalf} \& {Madau}}{{Metcalf} \&
  {Madau}}{2001}]{Metcalf2001}
{Metcalf} R.~B.,  {Madau} P.,  2001, \apj, 563, 9

\bibitem[\protect\citeauthoryear{{Miller}, {Heymans}, {Kitching}, {van
  Waerbeke}, {Erben}, {Hildebrandt}, {Hoekstra} \& {Mellier}}{{Miller}
  et~al.}{2013}]{Miller2013}
{Miller} L.,  {Heymans} C.,  {Kitching} T.~D.,  {van Waerbeke} L.,  {Erben} T.,
   {Hildebrandt} H.,  {Hoekstra} H.,    {Mellier} Y.,  2013, \mnras, 429, 2858

\bibitem[\protect\citeauthoryear{{Miller}, {Kitching}, {Heymans}, {Heavens} \&
  {van Waerbeke}}{{Miller} et~al.}{2007}]{Miller2007}
{Miller} L.,  {Kitching} T.~D.,  {Heymans} C.,  {Heavens} A.~F.,    {van
  Waerbeke} L.,  2007, \mnras, 382, 315

\bibitem[\protect\citeauthoryear{{Miyatake}, {More}, {Mandelbaum}, {Takada},
  {Spergel}, {Kneib}, {Schneider}, {Brinkmann} \& {Brownstein}}{{Miyatake}
  et~al.}{2013}]{Miyatake2013}
{Miyatake} H.,  {More} S.,  {Mandelbaum} R.,  {Takada} M.,  {Spergel} D.~N.,
  {Kneib} J.-P.,  {Schneider} D.~P.,  {Brinkmann} J.,    {Brownstein} J.~R.,
  2013, ArXiv e-prints

\bibitem[\protect\citeauthoryear{{Moster}, {Somerville}, {Maulbetsch}, {van den
  Bosch}, {Macci{\`o}}, {Naab} \& {Oser}}{{Moster} et~al.}{2010}]{Moster2010}
{Moster} B.~P.,  {Somerville} R.~S.,  {Maulbetsch} C.,  {van den Bosch} F.~C.,
  {Macci{\`o}} A.~V.,  {Naab} T.,    {Oser} L.,  2010, \apj, 710, 903

\bibitem[\protect\citeauthoryear{{Moustakas}, {Coil}, {Aird}, {Blanton},
  {Cool}, {Eisenstein}, {Mendez}, {Wong}, {Zhu} \& {Arnouts}}{{Moustakas}
  et~al.}{2013}]{Moustakas2013}
{Moustakas} J.,  {Coil} A.~L.,  {Aird} J.,  {Blanton} M.~R.,  {Cool} R.~J.,
  {Eisenstein} D.~J.,  {Mendez} A.~J.,  {Wong} K.~C.,  {Zhu} G.,    {Arnouts}
  S.,  2013, \apj, 767, 50

\bibitem[\protect\citeauthoryear{{Natarajan}, {De Lucia} \&
  {Springel}}{{Natarajan} et~al.}{2007}]{Natarajan2007}
{Natarajan} P.,  {De Lucia} G.,    {Springel} V.,  2007, \mnras, 376, 180

\bibitem[\protect\citeauthoryear{{Natarajan}, {Kneib}, {Smail}, {Treu},
  {Ellis}, {Moran}, {Limousin} \& {Czoske}}{{Natarajan}
  et~al.}{2009}]{Natarajan2009}
{Natarajan} P.,  {Kneib} J.-P.,  {Smail} I.,  {Treu} T.,  {Ellis} R.,  {Moran}
  S.,  {Limousin} M.,    {Czoske} O.,  2009, \apj, 693, 970

\bibitem[\protect\citeauthoryear{{Navarro}, {Frenk} \& {White}}{{Navarro}
  et~al.}{1997}]{NFW97}
{Navarro} J.~F.,  {Frenk} C.~S.,    {White} S.~D.~M.,  1997, \apj, 490, 493

\bibitem[\protect\citeauthoryear{{Neto}, {Gao}, {Bett}, {Cole}, {Navarro},
  {Frenk}, {White}, {Springel} \& {Jenkins}}{{Neto} et~al.}{2007}]{Neto2007}
{Neto} A.~F.,  {Gao} L.,  {Bett} P.,  {Cole} S.,  {Navarro} J.~F.,  {Frenk}
  C.~S.,  {White} S.~D.~M.,  {Springel} V.,    {Jenkins} A.,  2007, \mnras,
  381, 1450

\bibitem[\protect\citeauthoryear{{Nierenberg}, {Treu}, {Wright}, {Fassnacht} \&
  {Auger}}{{Nierenberg} et~al.}{2014}]{Nierenberg2014}
{Nierenberg} A.~M.,  {Treu} T.,  {Wright} S.~A.,  {Fassnacht} C.~D.,    {Auger}
  M.~W.,  2014, \mnras, 442, 2434

\bibitem[\protect\citeauthoryear{{Oguri} \& {Hamana}}{{Oguri} \&
  {Hamana}}{2011}]{Oguri2011}
{Oguri} M.,  {Hamana} T.,  2011, \mnras, 414, 1851

\bibitem[\protect\citeauthoryear{{Okabe}, {Futamase}, {Kajisawa} \&
  {Kuroshima}}{{Okabe} et~al.}{2014}]{Okabe2014}
{Okabe} N.,  {Futamase} T.,  {Kajisawa} M.,    {Kuroshima} R.,  2014, \apj,
  784, 90

\bibitem[\protect\citeauthoryear{{Pastor Mira}, {Hilbert}, {Hartlap} \&
  {Schneider}}{{Pastor Mira} et~al.}{2011}]{Pastor2011}
{Pastor Mira} E.,  {Hilbert} S.,  {Hartlap} J.,    {Schneider} P.,  2011, \aap,
  531, A169

\bibitem[\protect\citeauthoryear{{Peacock} \& {Smith}}{{Peacock} \&
  {Smith}}{2000}]{PeacockSmith2000}
{Peacock} J.~A.,  {Smith} R.~E.,  2000, \mnras, 318, 1144

\bibitem[\protect\citeauthoryear{{Rozo} \& {Rykoff}}{{Rozo} \&
  {Rykoff}}{2014}]{Rozo2014}
{Rozo} E.,  {Rykoff} E.~S.,  2014, \apj, 783, 80

\bibitem[\protect\citeauthoryear{{Rozo}, {Rykoff}, {Becker}, {Reddick} \&
  {Wechsler}}{{Rozo} et~al.}{2015}]{Rozo2015}
{Rozo} E.,  {Rykoff} E.~S.,  {Becker} M.,  {Reddick} R.~M.,    {Wechsler}
  R.~H.,  2015, \mnras, 453, 38

\bibitem[\protect\citeauthoryear{{Rykoff}, {Koester}, {Rozo}, {Annis},
  {Evrard}, {Hansen}, {Hao}, {Johnston}, {McKay} \& {Wechsler}}{{Rykoff}
  et~al.}{2012}]{Rykoff2012}
{Rykoff} E.~S.,  {Koester} B.~P.,  {Rozo} E.,  {Annis} J.,  {Evrard} A.~E.,
  {Hansen} S.~M.,  {Hao} J.,  {Johnston} D.~E.,  {McKay} T.~A.,    {Wechsler}
  R.~H.,  2012, \apj, 746, 178

\bibitem[\protect\citeauthoryear{{Rykoff}, {Rozo}, {Busha}, {Cunha},
  {Finoguenov}, {Evrard}, {Hao}, {Koester}, {Leauthaud}, {Nord}, {Pierre},
  {Reddick}, {Sadibekova}, {Sheldon} \& {Wechsler}}{{Rykoff}
  et~al.}{2014}]{Rykoff2014}
{Rykoff} E.~S.,  {Rozo} E.,  {Busha} M.~T.,  {Cunha} C.~E.,  {Finoguenov} A.,
  {Evrard} A.,  {Hao} J.,  {Koester} B.~P.,  {Leauthaud} A.,  {Nord} B.,
  {Pierre} M.,  {Reddick} R.,  {Sadibekova} T.,  {Sheldon} E.~S.,    {Wechsler}
  R.~H.,  2014, \apj, 785, 104

\bibitem[\protect\citeauthoryear{{Salpeter}}{{Salpeter}}{1955}]{Salpeter1955}
{Salpeter} E.~E.,  1955, \apj, 121, 161

\bibitem[\protect\citeauthoryear{{Shan}, {Kneib}, {Li}, {Comparat}, {Erben},
  {Makler}, {Moraes}, {Van Waerbeke}, {Taylor} \& {Charbonnier}}{{Shan}
  et~al.}{2015}]{Shan2015}
{Shan} H.,  {Kneib} J.-P.,  {Li} R.,  {Comparat} J.,  {Erben} T.,  {Makler} M.,
   {Moraes} B.,  {Van Waerbeke} L.,  {Taylor} J.~E.,    {Charbonnier} A.,
  2015, ArXiv e-prints

\bibitem[\protect\citeauthoryear{{Shirasaki}}{{Shirasaki}}{2015}]{Shirasaki2015}
{Shirasaki} M.,  2015, \apj, 799, 188

\bibitem[\protect\citeauthoryear{{Sif{\'o}n}, {Cacciato}, {Hoekstra},
  {Brouwer}, {van Uitert}, {Viola}, {Baldry} \& {Brough}}{{Sif{\'o}n}
  et~al.}{2015}]{Sifon2015}
{Sif{\'o}n} C.,  {Cacciato} M.,  {Hoekstra} H.,  {Brouwer} M.,  {van Uitert}
  E.,  {Viola} M.,  {Baldry} I.,    {Brough} S.,  2015, ArXiv e-prints

\bibitem[\protect\citeauthoryear{{Simha}, {Weinberg}, {Dav{\'e}}, {Fardal},
  {Katz} \& {Oppenheimer}}{{Simha} et~al.}{2012}]{Simha2012}
{Simha} V.,  {Weinberg} D.~H.,  {Dav{\'e}} R.,  {Fardal} M.,  {Katz} N.,
  {Oppenheimer} B.~D.,  2012, \mnras, 423, 3458

\bibitem[\protect\citeauthoryear{{Springel}, {Frenk} \& {White}}{{Springel}
  et~al.}{2006}]{Springel2006}
{Springel} V.,  {Frenk} C.~S.,    {White} S.~D.~M.,  2006, \nat, 440, 1137

\bibitem[\protect\citeauthoryear{{Springel}, {Wang}, {Vogelsberger}, {Ludlow},
  {Jenkins}, {Helmi}, {Navarro} \& {Frenk}}{{Springel}
  et~al.}{2008}]{springel2009}
{Springel} V.,  {Wang} J.,  {Vogelsberger} M.,  {Ludlow} A.,  {Jenkins} A.,
  {Helmi} A.,  {Navarro} J.~F.,    {Frenk} C.~S.,  2008, \mnras, 391, 1685

\bibitem[\protect\citeauthoryear{{Springel}, {White}, {Tormen} \&
  {Kauffmann}}{{Springel} et~al.}{2001}]{Springel2001}
{Springel} V.,  {White} S.~D.~M.,  {Tormen} G.,    {Kauffmann} G.,  2001,
  \mnras, 328, 726

\bibitem[\protect\citeauthoryear{{Taffoni}, {Mayer}, {Colpi} \&
  {Governato}}{{Taffoni} et~al.}{2003}]{Taffoni2003}
{Taffoni} G.,  {Mayer} L.,  {Colpi} M.,    {Governato} F.,  2003, \mnras, 341,
  434

\bibitem[\protect\citeauthoryear{{Tormen}, {Diaferio} \& {Syer}}{{Tormen}
  et~al.}{1998}]{Tormen1998}
{Tormen} G.,  {Diaferio} A.,    {Syer} D.,  1998, \mnras, 299, 728

\bibitem[\protect\citeauthoryear{{Vale} \& {Ostriker}}{{Vale} \&
  {Ostriker}}{2004}]{Vale2004}
{Vale} A.,  {Ostriker} J.~P.,  2004, \mnras, 353, 189

\bibitem[\protect\citeauthoryear{{Vegetti} \& {Koopmans}}{{Vegetti} \&
  {Koopmans}}{2009a}]{Vegetti2009a}
{Vegetti} S.,  {Koopmans} L.~V.~E.,  2009a, \mnras, 392, 945

\bibitem[\protect\citeauthoryear{{Vegetti} \& {Koopmans}}{{Vegetti} \&
  {Koopmans}}{2009b}]{Vegetti2009b}
{Vegetti} S.,  {Koopmans} L.~V.~E.,  2009b, \mnras, 400, 1583

\bibitem[\protect\citeauthoryear{{Vegetti}, {Koopmans}, {Bolton}, {Treu} \&
  {Gavazzi}}{{Vegetti} et~al.}{2010}]{Vegetti2010}
{Vegetti} S.,  {Koopmans} L.~V.~E.,  {Bolton} A.,  {Treu} T.,    {Gavazzi} R.,
  2010, \mnras, 408, 1969

\bibitem[\protect\citeauthoryear{{Vegetti}, {Lagattuta}, {McKean}, {Auger},
  {Fassnacht} \& {Koopmans}}{{Vegetti} et~al.}{2012}]{Vegetti2012}
{Vegetti} S.,  {Lagattuta} D.~J.,  {McKean} J.~P.,  {Auger} M.~W.,  {Fassnacht}
  C.~D.,    {Koopmans} L.~V.~E.,  2012, \nat, 481, 341

\bibitem[\protect\citeauthoryear{{Wang}, {Li}, {Kauffmann} \& {De
  Lucia}}{{Wang} et~al.}{2006}]{Wang2006}
{Wang} L.,  {Li} C.,  {Kauffmann} G.,    {De Lucia} G.,  2006, \mnras, 371, 537

\bibitem[\protect\citeauthoryear{{Wang}, {Li}, {Kauffmann} \& {De
  Lucia}}{{Wang} et~al.}{2007}]{Wang2007}
{Wang} L.,  {Li} C.,  {Kauffmann} G.,    {De Lucia} G.,  2007, \mnras, 377,
  1419

\bibitem[\protect\citeauthoryear{{Wetzel}, {Tinker}, {Conroy} \& {van den
  Bosch}}{{Wetzel} et~al.}{2013}]{Wetzel2013}
{Wetzel} A.~R.,  {Tinker} J.~L.,  {Conroy} C.,    {van den Bosch} F.~C.,  2013,
  \mnras, 432, 336

\bibitem[\protect\citeauthoryear{{Wright}, {Eisenhardt}, {Mainzer}, {Ressler},
  {Cutri}, {Jarrett} \& {Kirkpatrick}}{{Wright} et~al.}{2010}]{Wright2010}
{Wright} E.~L.,  {Eisenhardt} P.~R.~M.,  {Mainzer} A.~K.,  {Ressler} M.~E.,
  {Cutri} R.~M.,  {Jarrett} T.,    {Kirkpatrick} J.~D.,  2010, \aj, 140, 1868

\bibitem[\protect\citeauthoryear{{Xie} \& {Gao}}{{Xie} \&
  {Gao}}{2015}]{xie2015}
{Xie} L.,  {Gao} L.,  2015, \mnras, 454, 1697

\bibitem[\protect\citeauthoryear{{Xu}, {Mao}, {Wang}, {Springel}, {Gao},
  {White}, {Frenk}, {Jenkins}, {Li} \& {Navarro}}{{Xu} et~al.}{2009}]{xu2009}
{Xu} D.~D.,  {Mao} S.,  {Wang} J.,  {Springel} V.,  {Gao} L.,  {White}
  S.~D.~M.,  {Frenk} C.~S.,  {Jenkins} A.,  {Li} G.,    {Navarro} J.~F.,  2009,
  \mnras, 398, 1235

\bibitem[\protect\citeauthoryear{{Yang}, {Mo} \& {van den Bosch}}{{Yang}
  et~al.}{2003}]{Yang_etal2003}
{Yang} X.,  {Mo} H.~J.,    {van den Bosch} F.~C.,  2003, \mnras, 339, 1057

\bibitem[\protect\citeauthoryear{{Yang}, {Mo}, {van den Bosch}, {Jing},
  {Weinmann} \& {Meneghetti}}{{Yang} et~al.}{2006}]{Yang2006}
{Yang} X.,  {Mo} H.~J.,  {van den Bosch} F.~C.,  {Jing} Y.~P.,  {Weinmann}
  S.~M.,    {Meneghetti} M.,  2006, \mnras, 373, 1159

\bibitem[\protect\citeauthoryear{{Yang}, {Mo}, {van den Bosch}, {Pasquali},
  {Li} \& {Barden}}{{Yang} et~al.}{2007}]{Yang2007}
{Yang} X.,  {Mo} H.~J.,  {van den Bosch} F.~C.,  {Pasquali} A.,  {Li} C.,
  {Barden} M.,  2007, \apj, 671, 153

\bibitem[\protect\citeauthoryear{{Yang}, {Mo}, {van den Bosch}, {Zhang} \&
  {Han}}{{Yang} et~al.}{2012}]{Yang_etal2012}
{Yang} X.,  {Mo} H.~J.,  {van den Bosch} F.~C.,  {Zhang} Y.,    {Han} J.,
  2012, \apj, 752, 41

\bibitem[\protect\citeauthoryear{{Zhao}, {Jing}, {Mo} \& {B{\"o}rner}}{{Zhao}
  et~al.}{2003}]{Zhao2003}
{Zhao} D.~H.,  {Jing} Y.~P.,  {Mo} H.~J.,    {B{\"o}rner} G.,  2003, \apjl,
  597, L9

\bibitem[\protect\citeauthoryear{{Zhao}, {Jing}, {Mo} \& {B{\"o}rner}}{{Zhao}
  et~al.}{2009}]{Zhao2009}
{Zhao} D.~H.,  {Jing} Y.~P.,  {Mo} H.~J.,    {B{\"o}rner} G.,  2009, \apj, 707,
  354

\end{thebibliography}

\appendix

\section{Fake member contamination}

\begin{figure}
\includegraphics[width=0.5\textwidth]{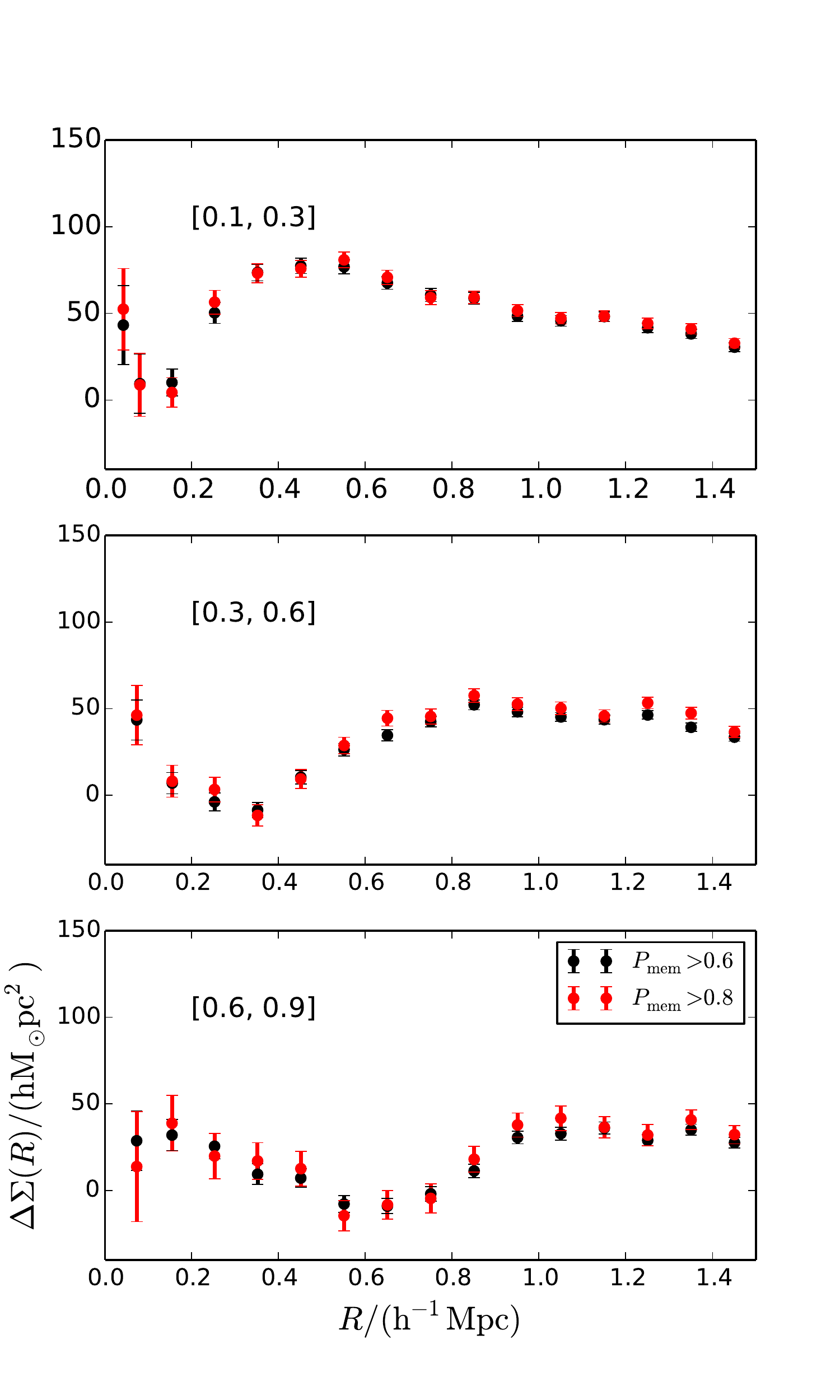}
\caption{Observed galaxy-galaxy lensing signal  for  satellite
  galaxies with different $P_{\rm mem}$ selection.}
  \label{fig:append}
\end{figure}

The uncertainties  of photometric redshifts cause the error in lens selection. The line-of-sight field galaxies may be misidentified as cluster members. These field galaxies are free of tidal disruption,  causing an overestimation of the $M_{\rm sub}/M_{\rm star}$ for satellite galaxies.  \citet{Rozo2015} investigate the cluster member identification, and find that the photometric 
membership estimator agrees with the spectroscopic membership with 1\% precision. In this work, we use satellite galaxy with membership probability larger than 80\%, which give us a mean probability of satellite $\sim80\%-90\%$. Thus, we conclude that
our satellite selection may face a  10\%-15\% contamination from the light sight field galaxies.  According to \citet{Li2013}, 10\% interlopers contribute to about 15\% of the total signal in the inner part.  The accurate 
estimation of the contamination requires a detailed comparison with mock catalog constructed in proper simulations, which is not available at the moment. We thus made the following test to explore the effects of the interlopers. Instead of choosing satellite galaxies with $P_{\rm mem}>80\%$, we choose satellite galaxies with $P_{\rm mem}>60\%$, and re-calculate the lensing signal. 
We show the comparison of the lensing signal in Fig.\ref{fig:append}. The later $P_{\rm mem}$ selection will increase the interloper fraction from 10\%-15\% to 20\%-25\%. We find that the change in the lensing is only about 10\% at the subhalo dominating region, smaller than the statistical error bars.

\section{Photometric redshift bias test}

\begin{figure}
\includegraphics[width=0.6\textwidth]{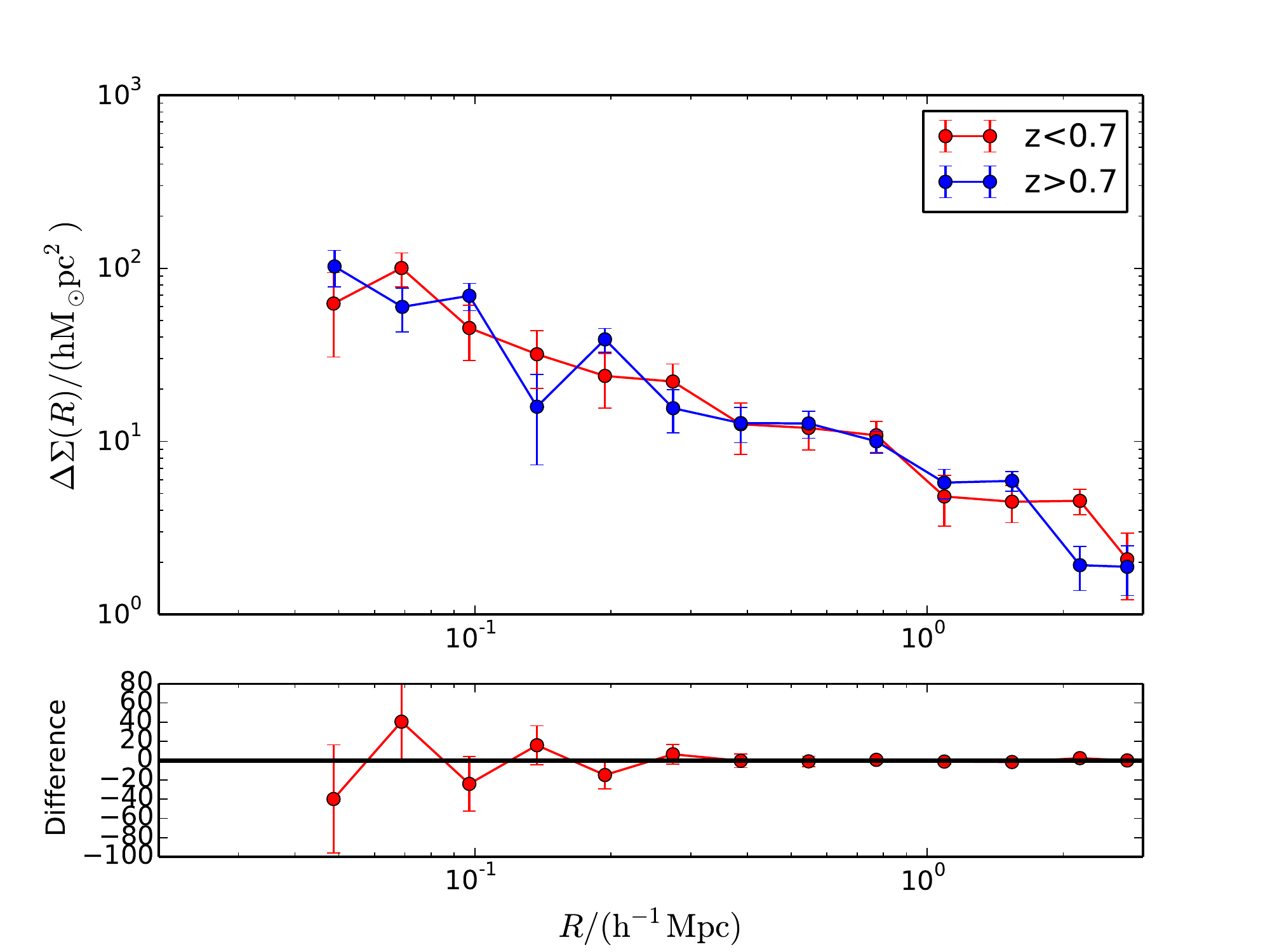}
\caption{Observed galaxy-galaxy lensing signal  with high-z sources  (blue) and 
 low-z  sources (red). The lenses are selected from the LOWZ and CMASS galaxies. The lower panel shows the difference 
 between results of the two sample of sources. }
 \label{fig:append2}
\end{figure}

The biased photometric redshifts can cause a systematic error in the measured lensing signal. We make the following test to explore this uncertainty. We split the source galaxy sample in two, the high-z ($z>0.7$) and the low-z ($z<0.7$) samples. We then compute $\Delta\Sigma(R)$ for the same set of lenses with each half of the sources.  We fit $\Delta\Sigma_{high-z} = (1+b)\Delta\Sigma_{\rm low-z}$ where b is a free parameter, which is used to estimate the bias introduced by photo-z error. If the source redshifts are free from bias, we will have b=0. In this test, the lens sample can be any galaxies as long as their redshift distribution is the same with that of our satellite samples in the paper. To maximum the lensing signal, we thus select galaxies from the LOWZ and CMASS galaxies from SDSS-BOSS survey\citep{Dawson2013}, and require them to have the same redshift distribution as the satellite galaxies in our paper.  The results are shown in Fig.\ref{fig:append2}. We find that the b parameter is consistent with 0 within the statistical errors ($b= -0.05\pm0.19$). 

\end{document}